\documentclass[11pt]{article}
\usepackage[utf8]{inputenc}
\usepackage{amssymb,amsmath}
\usepackage{bm} 
\usepackage{booktabs} 
\usepackage{array}
\usepackage{latexsym}
\usepackage{graphicx}
\usepackage{color}
\usepackage{datetime}
\usepackage[nosort]{cite}
\usepackage{verbatim}
\usepackage{enumerate}
\usepackage{chngpage} 
\usepackage{mathrsfs}
\usepackage{euscript}
\usepackage{psfrag}

\usepackage{datetime}

\usepackage[nosort]{cite}
\usepackage{chngpage} 
\usepackage{setspace}
\usepackage{tensor}
\usepackage{physics}

\usepackage{mciteplus}

\usepackage[colorlinks=true,      linkcolor=blue,      urlcolor=blue,      
            filecolor=blue,      citecolor=blue,       pdfstartview=FitH,     
						pdfpagemode=UseNone,      bookmarksopen=true]{hyperref}  
\usepackage[all]{hypcap}     

\definecolor{cardinal}{rgb}{0.6,0,0}
\definecolor{darkgreen}{rgb}{0,0.4,0}
\definecolor{golden}{rgb}{0.92, 0.7, 0}
\definecolor{midnight}{rgb}{0, 0, 0.5}
\definecolor{darkblue}{rgb}{0, 0, 0.7}
\definecolor{purple}{rgb}{0.5, 0, 0.5}

\def\oneone{\rlap 1\mkern4mu{\rm l}}

\def\coeff#1#2{\relax{\textstyle \frac{#1}{#2}}\displaystyle}

\def\IR{\mathbb{R}}

\def\cL{{\cal L}}

\def\nBPS#1{$\frac{1}{#1}$-BPS}

\def\mname{maze}

\numberwithin{equation}{section}

\begin{document}

\phantom{AAA}
\vspace{-10mm}

\begin{flushright}
%
%
\end{flushright}

\vspace{1.9cm}

\begin{center}

{\huge {\bf   Maze Topiary in Supergravity }}\\

{\huge {\bf \vspace*{.25cm}  }}

\vspace{1cm}

{\large{\bf { Iosif Bena$^{1}$, Anthony Houppe$^{2}$, Dimitrios Toulikas$^{1}$   \\ and  Nicholas P. Warner$^{1,3,4}$}}}

\vspace{1cm}

$^1$Institut de Physique Th\'eorique, \\
Universit\'e Paris Saclay, CEA, CNRS,\\
Orme des Merisiers, Gif sur Yvette, 91191 CEDEX, France \\[12pt]

$^2$Institut f\"ur Theoretische Physik, ETH Z\"urich, \\
Wolfgang-Pauli-Strasse 27, 8093 Z\"urich, Switzerland \\[12pt]

\centerline{$^3$Department of Physics and Astronomy}
\centerline{and $^4$Department of Mathematics,}
\centerline{University of Southern California,} 
\centerline{Los Angeles, CA 90089, USA}

\vspace{10mm} 
{\footnotesize\upshape\ttfamily iosif.bena @ ipht.fr,  ahouppe @ phys.ethz.ch, dimitrios.toulikas @ ipht.fr,  warner @ usc.edu} \\

\vspace{2.2cm}
 
\textsc{Abstract}

\end{center}

\begin{adjustwidth}{3mm}{3mm} 
 
We show that the supergravity solutions for  \nBPS{4} intersecting systems of M2 and M5 branes are completely characterized by a single ``\mname'' function that satisfies a non-linear  ``\mname'' equation similar to the Monge-Amp\`ere equation. We also show that the near-brane limit of certain intersections are AdS$_3 \times$S$^3 \times$S$^3$ solutions warped over a Riemann surface, $\Sigma$.  There is an extensive literature on these subjects and we construct  mappings between various approaches and use brane probes to elucidate the relationships between the M2-M5 and AdS systems.  We also use dualities to map our results onto other systems of intersecting branes.  This work is motivated by the recent realization that adding momentum to M2-M5 intersections gives a supermaze that can reproduce the black-hole entropy without ever developing an event horizon. We take a step in this direction by adding a certain type of momentum charges that blackens the M2-M5 intersecting branes. The near-brane limit of these solutions is a BTZ$^{\rm extremal} \times$S$^3 \times$S$^3 \times \Sigma$ geometry in which the BTZ  momentum is a function of the Riemann surface coordinates.

\vspace{-1.2mm}
\noindent

\end{adjustwidth}

\thispagestyle{empty}
\newpage


\tableofcontents

\section{Introduction}
\label{sec:Intro}

The entropy of many classes of brane systems  can be counted using perturbative String Theory in a regime of parameters in which gravity is turned off. The result matches the entropy of the black hole with the same charges in the regime of parameters in which gravity is turned on. This gives spectacular matches, both for the D1-D5 system \cite{Sen:1995in,Strominger:1996sh}, for M5-M5-M5-P black holes \cite{Maldacena:1997de}, and also for Type IIA F1-NS5-P black holes \cite{Dijkgraaf:1996cv}.

These entropy-matching computations rest on the fact that the counting of index states  essentially does not change\footnote{There can be jumps under ``wall-crossing'' but these are sub-leading.}  as couplings are varied from perturbative string states to black-hole microstructure.  Such an approach fails to address  the hugely important issue of  what happens to particular individual microstates as one turns gravity on, and precisely what the microstate structure ``looks like'' at finite $G_{\rm Newton}$?  An alternative formulation of this question is: what distinguishes different black-hole microstates from each other in the regime of parameters where the classical black hole exists. There are strong arguments, coming from  quantum information theory,  that individual microstates should differ from each other and from the classical black-hole solution at the scale of the horizon \cite{Mathur:2009hf,Almheiri:2012rt}.    Indeed, several very large classes of microstate geometry solutions, dual to some families of pure states of the CFT that counts the black hole entropy, have been constructed, both for supersymmetric black holes \cite{Shigemori:2013lta,Giusto:2013bda,Bena:2015bea,Bena:2016ypk,Bena:2017geu,Bena:2017upb,Bena:2017xbt,Ceplak:2018pws,Heidmann:2019xrd,Heidmann:2019zws,Walker:2019ntz,Ganchev:2021iwy,Bena:2022sge,Ceplak:2022wri,Ganchev:2022exf,Ceplak:2022pep}, and also, in fundamentally different approaches, for non-extremal ones \cite{Ganchev:2021pgs,Ganchev:2021ewa,Ganchev:2023sth} and \cite{Jejjala:2005yu, Bena:2009qv, Bobev:2011kk, Vasilakis:2011ki, Bena:2015drs,Bena:2016dbw,Bossard:2017vii,Bah:2020ogh,Bah:2020pdz,Bah:2021owp,Bah:2021irr,Bah:2022yji,Bah:2022pdn,Bah:2023ows}.

Tracking the D1-D5 microstates from weak to strong coupling is challenging, since the momentum is carried by bi-fundamental strings whose back-reaction is only known at the most rudimentary level.  The construction of superstrata \cite{Shigemori:2013lta,Giusto:2013bda,Bena:2015bea,Bena:2016ypk,Bena:2017geu,Bena:2017upb,Bena:2017xbt,Ceplak:2018pws,Heidmann:2019xrd,Heidmann:2019zws,Walker:2019ntz,Ganchev:2021iwy,Bena:2022sge,Ceplak:2022wri,Ganchev:2022exf,Ceplak:2022pep} largely rests on collective string excitations in the untwisted sectors of the D1-D5 CFT. While these solutions describe a significant sector of the black-hole microstructure, they fall parametrically short of capturing the black-hole entropy \cite{Shigemori:2019orj,Mayerson:2020acj}.  To obtain a geometric description of  generic microstructure one must capture coherent combinations of  twisted-sector states  of the CFT, and this  seems to be easier in the Type IIA F1-NS5-P formulation of the brane system that leads to a black hole. 

In this formulation, the momentum is carried by little strings, which live on the NS5 world-volume. These little strings have a very simple geometric description: When uplifting the F1-NS5 system to M theory, each F1 uplifts to an M2 wrapping the M-theory direction. This M2 can break into $N_5$ strips (which correspond to the little strings on the NS5 world-volume) which carry momentum independently. The $N_1 N_5$ resulting little strings form a complicated maze of intersecting M2 and M5 branes carrying momentum and whose entropy (upon taking into consideration fermionic partners) matches exactly that of the F1-NS5-P black hole. The beauty of this characterization of the microstructure is that it lends itself to a geometric description of  the coherent states in terms of supergravity.  

Since the momentum of such a  ``supermaze'' is carried by waves on the little strings, the microstates of the black hole have coherent expression as momentum carried by components of a fractionated M2-M5 system.  One can thus explore such structures in the regime of parameters where gravity is large and the classical black hole solution is valid. As we have seen in \cite{Bena:2022wpl}, upon taking into account brane-brane interaction, the supermaze has 16 supercharges locally, but only 4 supercharges globally. This is a property shared by all brane systems whose supergravity back-reaction gives a smooth horizonless solution \cite{Bena:2022fzf}, which makes us confident that the fully back-reacted supergravity solution sourced by the supermaze will not have a horizon. If the supergravity formulation of the supermaze turns out as we expect it will, it would finally provide a proof of the fuzzball conjecture.

The purpose of this paper is to make a crucial first step in developing the supergravity formulation of the supermaze.   As one would expect, the supergravity solutions for generic intersecting branes are extremely  complicated.  Moreover,  supergravity solutions for various intersecting branes have been extensively studied in the past. We start by pulling together, and unifying, earlier literature on the  intersecting-brane solutions relevant to the supermaze.    We obtain the supergravity equations governing supermaze solutions, and show how a ``near-brane'' limit is related to a certain class of warped AdS$_3 \times$S$^3 \times$S$^3$ solutions constructed in \cite{Bachas:2013vza}. 

There are several stages in this construction, and several technical tools we will develop. The first is to construct the solution corresponding to the supermaze without momentum. We will do this from first principles in Section \ref{sec:GenRes}, and relate our equations and solutions to the construction in \cite{Lunin:2007mj}.  We also show that, if one imposes spherical ($SO(4) \times SO(4)$) symmetry, then our equations capture all the \nBPS{4} M2-M5 solutions.   This is described in Section \ref{ss:sphsymm} and Appendix \ref{app:SphericalSols}. Even without spherical symmetry, the results of \cite{Lunin:2007mj}  suggest that the results described in Section \ref{sec:GenRes} capture all the possible  pure intersecting M2-M5 solutions.

There is an important issue that we clarify in Appendix \ref{app:democracy}.  We are considering the \nBPS{4} system (8 supersymmetries) of intersecting M2's and M5's. These branes have one spatial direction in common, which we label by $y$.   The combined M2 and M5 system therefore spans six spatial dimensions, and so has four transverse spatial dimensions.  Because of the way that the supersymmetry projectors work, one can add, {\it without breaking the supersymmetry any further}, a complementary set of M5 branes, which we will denote as M5', whose world volume spans these four transverse dimensions and $y$.  There is a  complete democracy between the original M5  branes and the M5' branes.   One can thus have \nBPS{4}  solutions with arbitrary numbers of M2, M5 and M5' branes, and the BPS equations respect this fact. However, the explicit eleven-dimensional metric involves a fibration that seemingly breaks the democracy.  In Appendix \ref{app:democracy} we discuss how this seeming asymmetry between the M5  branes and the M5' branes is simply an artifact of coordinate choices.  

From the perspective of the supermaze, we want the M5 branes to wrap what will become  compactified directions and {\it not fill the space-time}.  We thus focus on solutions with no net M5' charge.  However, because we do want net M2 and M5 charges, the Chern-Simons term of supergravity will generically require some, at least,  ``dipolar'' distribution of M5' charge.  These considerations play a major role in determining the solutions we consider in subsequent sections.

In Section  \ref{sec:AdSlmits}, we first look at a smeared, highly symmetric version of our supermaze and show how it is related to a brane-intersection solution found in  \cite{deBoer:1999gea}.  We then consider a  more general scaling limit of our system of equations that corresponds to a ``near-brane-intersection'' limit of the supermaze.  We show that this reduces to a particular family\footnote{The fact that our system asymptotes to M2 or M5 branes at infinity implies that we must set   $\gamma =1$ in the  solutions constructed in \cite{Bachas:2013vza}.} of the AdS$_3 \times$S$^3 \times$S$^3$ solutions constructed in \cite{Bachas:2013vza}.   Our analysis provides the complete mapping between the near-brane supermaze, the results of \cite{Lunin:2007mj} and the results of \cite{Bachas:2013vza}. We re-derive the BPS equations of \cite{Bachas:2013vza} from the perspective of the supermaze, thereby furnishing a description of the supersymmetries of the near-brane, AdS formulation in terms of projection matrices in M-theory. 

In Section \ref{sec:stringwebs} and Appendix \ref{app:duality-D2-D4}, we   show how our supermaze system  can be smeared and dualized into various  brane systems.    In particular,  we show how the supermaze solutions can be dualized to the F1-D1 string web, whose geometry was constructed in \cite{Lunin:2008tf}.  In Appendix  \ref{app:tiltedD2},  we also use dualities to construct simple, new solutions to our original supermaze equations. 

In Section \ref{sec:floatingM2} we consider ``floating'' M2 and M5 branes both in the original intersecting M2-M5 brane formulation and in the near-brane AdS formulation. Floating branes \cite{Bena:2009fi} reveal the probes that are mutually BPS with respect to the background brane configuration. While the floating brane analysis is relatively straightforward in the M2-M5 formulation, it is particularly revealing in the AdS formulation of \cite{Bachas:2013vza}.  Indeed it shows how the AdS directions emerge from combinations of natural brane coordinates and shows that only a particular family of the solutions considered in \cite{Bachas:2013vza} correspond to brane configurations that are asymptotic to M2 or M5 branes at infinity.

In Section \ref{sec:examples}, we adapt and develop some of the examples of AdS solutions obtained in \cite{Bachas:2013vza}, mapping them across to the M2-M5 brane-intersection formulation.  This reveals how the AdS space and the Riemann surface of \cite{Bachas:2013vza} appear in  the more intuitive brane configurations that are inherent to the supermaze.

The primary core of this paper is the development of ``momentum-free'' supermazes, the equations that govern them and how to map the near-brane, AdS solutions onto the M2-M5 configurations.  The next step in this program will be to add independent momenta to all the elements of this system.  This is going to be a challenging enterprise for future work.  However,  we could not resist exploring the addition of a simple momentum charge as a first step in this direction.  In Section \ref{sec:Mom} we show that a singular momentum charge can indeed be added by a harmonic Ansatz for the distribution of BPS momentum charge.  On top of  the AdS, near-brane  ``momentum-free'' supermaze, adding such a momentum distribution converts the AdS$_3$ factor into an extremal spinning BTZ black-hole geometry whose momentum charge depends on the  two dimensions of the brane intersection locus. The fact that adding such a momentum charge involves such an extremely simple Ansatz makes us very optimistic about completing the far more ambitious project of adding independent  momenta to each intersection locus. Even if the asymptotics of these solutions is not flat, it is also worth remarking that they give an infinite violation of black-hole uniqueness in this system.

We finish by making some concluding remarks in Section \ref{sec:Concl}.

\section{The most general solution describing M5-M2 intersections}
\label{sec:GenRes}


We are interested in 8-supercharge, or \nBPS{4}, supergravity solutions describing the uplift of momentum-free type IIA little strings inside NS5 branes. If we denote the direction of the little strings as $x^1$, and the M-theory direction as $x^2$, the M-theory solution will have the charges of M2 branes extended along $012$ and of M5 branes extended along the directions $013456$. Before the back-reaction of the branes, one can think about this configuration as describing M5 branes located at arbitrary positions in the M-theory direction, $x^2$, and M2 branes stretched between any of these M5 branes, and located at arbitrary locations inside the four-torus spanned by $x^3,x^4,x^5,x^6$. 

However, we know that this picture is altered by the interaction between these branes \cite{Bena:2022wpl}:  the M2 branes will pull on the M5 branes, and the final brane configuration will consist of multiple spikes with M5 and M2 charge, extending from one M5 to another. Furthermore, we expect the back-reaction of these spikes to give rise, via a geometric transition,  to a new geometry containing bubbles and fluxes, but no brane sources. However, both the brane interactions and the geometric transition will respect the symmetries and the supersymmetries of the original brane system. 

Our strategy is to use the eight Killing spinors of the system,  defined in terms of the frame components along the M2 and M5 directions: 
\begin{equation}
 \Gamma^{012} \, \varepsilon  ~=~ - \varepsilon \,,   \qquad  \Gamma^{013456} \, \varepsilon  ~=~\varepsilon 
 \label{projs1}
\end{equation}
to solve the gravitino equation 
\begin{equation}
\delta \psi_\mu ~\equiv~ \nabla_\mu \, \epsilon ~+~ \frac{1}{288}\,
\Big({\Gamma_\mu}^{\nu \rho \lambda \sigma} ~-~ 8\, \delta_\mu^\nu  \, 
\Gamma^{\rho \lambda \sigma} \Big)\, F_{\nu \rho \lambda \sigma}\, \epsilon ~=~ 0 \,,
\label{11dgravvar}
\end{equation}
and to determine the metric and three-form vector potential of this system. Before beginning we can observe that, since 
\begin{equation}
 \Gamma^{0123456789\,10}    ~=~ \oneone \,,  
 \label{prodgammas}
\end{equation}
equation  \eqref{projs1}  implies that
\begin{equation}
 \Gamma^{01789\,10} \, \varepsilon  ~=~-\varepsilon \,,
 \label{projs2}
\end{equation}
and hence adding a set of M5 branes along $01789\,10$ does not break supersymmetry any further.  We will denote this second possible set of branes by M5'.

\subsection{The metric and the three-form potential.}
\label{ss:metric}

We parametrize the M2 directions via $(x^0, x^1,x^2) = (t,y,z)$, and we denote the coordinates inside the M5 branes $(x^3, \dots,x^6)$  by  vectors $\vec u \in \IR^4$.  The transverse dimensions, $(x^7, \dots,x^{10})$, will be parametrized by vectors  $\vec v \in {\IR}^4$.  As we explain in Appendix \ref{app:SphericalSols}, upon using \eqref{11dgravvar} and the equations of motion of eleven-dimensional supergravity we find that the eleven-dimensional metric ultimately has the form:
\begin{equation}
\begin{aligned}
ds_{11}^2 ~=~  e^{2  A_0}\, \Big[ -dt^2 &~+~ dy^2 ~+~ e^{-3  A_0} \, (-\partial_z w )^{-\frac{1}{2}}\, d \vec u \cdot d \vec u  ~+~ e^{-3  A_0} \, (-\partial_z w )^{\frac{1}{2}}\, d \vec v \cdot d \vec v \,    \\
  &  ~+~  (-\partial_z w ) \, \big( dz ~+~(\partial_z w )^{-1}\,   (\vec \nabla_{\vec u} \, w )  \cdot  d \vec u \big)^2  \Big]\,.
\end{aligned}
 \label{11metric}
\end{equation}
This metric is conformally flat along time and the common M2-M5 direction, $(t,y) \in \IR^{(1,1)}$, and also along the internal M5 torus (parameterized by  $\vec u)$ and the transverse $\IR^4$ parameterized by $\vec v \in \IR^4$. Since the equations we solve are local, the torus wrapped by the M5 branes can be replaced by $\IR^4$. To  obtain solutions with a compact four-torus one  has to consider periodic sources in this $\IR^4$. 
The metric involves a non-trivial fibration of the ``M-theory direction,'' $z$, over this internal  $\IR^4$.  

The  constraints on, and relationships between, the functions $A_0(\vec u, \vec v, z)$ and $w(\vec u, \vec v, z)$ will be discussed below, and, for obvious reasons, we require $\partial_z w <0$.

We will use the set of frames:
\begin{equation}
\begin{aligned}
e^0 ~=~  & e^{A_0}\, dt \,, \qquad e^1~=~  e^{A_0}\, dy \,,  \qquad e^2 ~=~   e^{A_0} (-\partial_z w )^{\frac{1}{2}} \, \Big( dz ~+~(\partial_z w )^{-1}\,  \big (\vec \nabla_{\vec u} \, w \big)  \cdot  d \vec u \Big) \,, \\
e^{i+2}~=~  & e^{- \frac{1}{2} A_0} \, (-\partial_z w )^{-\frac{1}{4}}\,  du_i \,, \qquad e^{i+6} ~=~  e^{- \frac{1}{2} A_0} \, (-\partial_z w )^{\frac{1}{4}}\,  dv_i\,,   \qquad {i = 1,2,3,4}   \,.
\end{aligned}
 \label{11frames}
\end{equation}
The three-form vector potential is given by:
\begin{equation}
C^{(3)} ~=~   - e^0 \wedge e^1 \wedge e^2 ~+~ \frac{1}{3!}\, \epsilon_{ijk\ell} \,  \big((\partial_z w )^{-1}\, (\partial_{u_\ell} w) \,  du^i \wedge du^j \wedge du^k ~-~ (\partial_{v_\ell} w)  \, dv^i \wedge dv^j \wedge dv^k  \big)  \,,
 \label{C3gen}
\end{equation}
where $\epsilon_{ijk\ell}$ is the $\epsilon$-symbol on $\IR^4$. 
 
This solution appears to be asymmetric between the two $\IR^4$'s, and hence between the M5 and M5' branes. However, as we explain in detail in Appendix \ref{app:democracy}, this is a coordinate artifact coming from the choice of fibration of the M-theory direction. One can flip the fibration from the $\vec u$-plane to the  $\vec v$-plane by using $w$ as a coordinate and thinking of $z$ as a function, $z(w,\vec u,\vec v)$.  

\subsection{The \mname\ function}
\label{ss:master}

Denote the Laplacians on each $\IR^4$ via:
\begin{equation}
{\cal L}_{u} ~\equiv~  \nabla_{\vec u} \cdot \nabla_{\vec u} \,, \qquad {\cal L}_{v} ~\equiv~  \nabla_{\vec v} \cdot \nabla_{\vec v} \,,
 \label{Laps}
\end{equation}
and suppose that $G_0(\vec u, \vec v, z)$ is a solution to what we will refer to  as the ``\mname\ equation\footnote{In other contexts, when a solution to a BPS system is governed by a single function satisfying one equation, this function and the equation have been referred to as a ``master function''  and a ``master equation.'' (See, for example, \cite{Pilch:2004yg,Nemeschansky:2004yh,Bena:2004jw}.)  Since $G_0$ completely encodes the structure of the ``maze'' of branes, we think ``\mname'' is a more appropriate sobriquet here.}:''
\begin{equation}
 {\cal L}_{v} G_0 ~=~  ({\cal L}_{u} G_0)\,(\partial_z \partial_z G_0) ~-~ ( \nabla_{\vec u} \partial_z G_0)\cdot  (  \nabla_{\vec u} \partial_z G_0)\,.
 \label{master}
\end{equation}
One then finds that there are eight solutions to the gravitino variation equations, (\ref{11dgravvar}), provided $w$ and $A_0$ are given by:
\begin{equation}
w ~=~ \partial_z G_0  \,, \qquad  e^{-3  A_0} \, (-\partial_z w )^{\frac{1}{2}} ~=~ {\cal L}_{v}  G_0 \,.
 \label{solfns}
\end{equation}
One can also verify that these equations along with \eqref{master} imply
\begin{equation}
 e^{-3  A_0} \, (\partial_z w )^{-\frac{1}{2}} ~-~ (\partial_z w )^{-1} \, (\nabla_{\vec u} \,w )\cdot  (\nabla_{\vec u} \, w ) ~=~  - {\cal L}_{u}  G_0  \,.
 \label{reln1}
\end{equation}

Hence ``brane-intersection'' equations, like  \eqref{master}, should determine the M5-M2 intersections of interest to us.  More precisely, one expects that once the  boundary conditions and sources are specified,  equation \eqref{master} should have a unique solution.

The differential equation \eqref{master} has a very interesting form but it is non-linear  and cannot be explicitly solved in general, and rigorous existence proofs are extremely challenging.  (It also has variant, but very similar, forms for many other solutions describing \nBPS{4} brane intersections \cite{Lunin:2007ab,Lunin:2007mj,Lunin:2008tf}).   Nevertheless, it was argued in \cite{Lunin:2007mj} (see, sections 4.5 and 5.1) using perturbation theory that once one has specified a  brane distribution through its  boundary conditions and sources, there is indeed a unique solution to \eqref{master}, and thus there is a one-to-one map between brane webs\footnote{Some of these brane webs are a special examples of the configurations we consider, where one smears over three directions of the internal four-torus.} and solutions to \eqref{master}.

\subsection{Imposing spherical symmetry}
\label{ss:sphsymm}

A  supermaze generically has spherical symmetry in the transverse $\IR^4$, but breaks all isometries in the internal $\IR^4$. Since this solution is complicated, one can try focusing on a simpler solution that has spherical symmetry in the internal $\IR^4$ as well. This solution can describe either a single M2 spike ending on and pulling on  an M5 brane, or an M2 brane stretching between two M5 branes, or multiple coincident M2 branes ending on multiple M5 branes.

The metric with spherical symmetry in the two $\IR^4$'s is:
\begin{equation}
\begin{aligned}
ds_{11}^2 ~=~ e^{2  A_0}\, \Big[&  - dt^2 ~+~ dy^2  ~+~  (-\partial_z w ) \, \big( dz ~+~(\partial_z w )^{-1}\,  (\partial_u w)  \, d u \big)^2   \\
& ~+~ e^{-3  A_0} \, (-\partial_z w )^{-\frac{1}{2}}\, \big( du^2  ~+~ u^2 \, d\Omega_3^2 \big)   ~+~ e^{-3  A_0} \, (-\partial_z w )^{\frac{1}{2}}\, \big( dv^2  ~+~ v^2 \, d{\Omega'}_3^2 \big)  
  \Big]\,,
\end{aligned}
 \label{11metric-symm}
\end{equation}
where $u = |\vec u|$, $v = |\vec v|$ and $d\Omega_3^2$, $d{\Omega'}_3^2$ are the metrics of unit three-spheres in each $\IR^4$ factor. 
The obvious choice for a set of frames is then:
\begin{equation}
\begin{aligned}
e^0 ~=~  &  e^{A_0}\, dt \,, \qquad e^1~=~  e^{A_0}\,  dy \,,  \qquad e^2 ~=~  e^{A_0}\,   (-\partial_z w )^{\frac{1}{2}} \,  \big( dz ~+~(\partial_z w )^{-1}\,   (\partial_u w )  \, d u \big)  \,, \\  e^3 ~=~ &   e^{- \frac{1}{2} A_0} \, (-\partial_z w )^{-\frac{1}{4}}\, du \,,\qquad e^4~=~  e^{- \frac{1}{2} A_0} \, (-\partial_z w )^{\frac{1}{4}}\,  \, dv  \,,\\
e^{i+4}~=~& e^{- \frac{1}{2} A_0} \, (-\partial_z w )^{-\frac{1}{4}}\,    \sigma_i  \,, \qquad e^{i+7}  ~=~ e^{- \frac{1}{2} A_0} \, (-\partial_z w )^{\frac{1}{4}}\, \tilde \sigma_i \,,   \qquad {i = 1,2,3}   \,.
\end{aligned}
 \label{firstframes}
\end{equation}
where $\sigma_i$ and $\tilde \sigma_i$ are left-invariant one-forms on the unit three-spheres.

Similarly one has the spherically symmetric $3$-form potential:
\begin{equation}
C^{(3)} ~=~   - e^0 \wedge e^1 \wedge e^2 ~+~  (\partial_z w )^{-1} \, \big (u^3 \partial_u w \big)  \, {\rm Vol}({S^3}) ~+~  \, \big(v^3 \partial_v w \big) \,  {\rm Vol}({{S'}^3})    \,,
 \label{C3symm}
\end{equation}
where ${\rm Vol}({S^3})$ and ${\rm Vol}({{S'}^3})$ are the volume forms of the unit three-spheres.  Note there is a sign-flip of the flux along the ${S'}^3$ compared to  (\ref{C3gen}).  This is because of the orientation change in  (\ref{firstframes})  compared to (\ref{11frames}) where the $e^4$ is now the radial $v$-direction.

The spherically symmetric formulation is important because it is the one we use most directly, and because it is relatively easily to show that it is the most general \nBPS{4} configuration for our intersecting M2 and M5 branes.

In Appendix \ref{app:SphericalSols} we derive this solution following the methodology developed in \cite{Gowdigere:2002uk,Pilch:2003jg,Gowdigere:2003jf,Nemeschansky:2004yh,Pilch:2004yg}. We will show that  the solutions described above are the only possible ones with these symmetries.

\section{Near-brane M5-M2 intersections}
\label{sec:AdSlmits}

Perhaps rather surprisingly, the  \nBPS{4} geometry  created by intersecting M2 and M5 branes has a  near-brane limit that includes an AdS$_3$  factor.  One can see this by searching for solutions with an $SO(2,2) \times SO(4) \times SO(4)$ isometry and whose geometry contains factors of AdS$_3 \times$S$^3 \times$S$^3$. The most general such geometry can depend on two non-trivial variables that we will label as $(\rho, \xi)$.  Such  solutions have been extensively studied in  \cite{deBoer:1999gea,Lunin:2007ab,DHoker:2008lup,DHoker:2008rje,DHoker:2008wvd,DHoker:2009lky, DHoker:2009wlx,Bachas:2013vza}. 

\subsection{Smeared solutions}
\label{ss:smeared}

One can smear along the M-theory direction and  thereby find geometries that are ultimately independent of $z$.  This results in the solution given in \cite{deBoer:1999gea}.  However one has to be a little careful in using the methodology of Section \ref{sec:GenRes} to arrive at this result:  smearing should make the solution independent of the M-theory direction but, as we will describe, this requires a judicious coordinate change. 

There are two ways to proceed.  The smearing will wash out the fibration and so one can re-work the approach of Appendix  \ref{app:SphericalSols} but starting with $B_1 \equiv 0$. 

 In this instance one finds that the BPS equations only solve a subset of the equations of motion and so one must supplement the BPS system with one of the equations of motion.   Alternatively, one can use the results of Section \ref{sec:GenRes} while being careful about what it means to be independent of the M-theory direction.  Specifically, we will see that  to realize such independence one may have to change the $z$-coordinate via $\hat z =  z f(u)$ to get a metric that is then independent of $\hat z$.  In particular, such a coordinate change leads to a differential $d \hat z =f(u) d z + z f'(u) du$ that can be used to absorb a $z$-dependent   $B_1$ field into a coordinate re-definition. 

To explore these possibilities, and cast the net a little wider, it is instructive  to seek solutions  to (\ref{master}) with a power-series Ansatz in $z$: 
\begin{equation}
G_0  ~=~  - \frac{1}{2} \, z^2 \, \hat g_2(u,v)  ~+~  z \, \hat g_1(u,v) ~+~ \hat g_0(u,v) \,.
 \label{G0simp}
\end{equation}
Substituting this into (\ref{master}) results in a quadratic in $z$ and hence three equations:
\begin{equation}
\begin{aligned}
&\cL_{\vec v} \, \hat g_2 ~+~ \hat g_2 \, \cL_{\vec u} \, \hat g_2 ~-~ 2 \, \big(\vec \nabla_{\vec u}  \, \hat g_2\big)^2    ~=~ 0 \,, \\ 
&\cL_{\vec v} \, \hat g_1 ~+~ \hat g_2 \, \cL_{\vec u}  \,\hat g_1 ~-~ 2 \, \big(\vec \nabla_{\vec u} \, \hat g_1\big)\cdot  \big(\vec \nabla_{\vec u} \, \hat g_2\big)      ~=~ 0  \,, \\
&\cL_{\vec v} \, \hat g_0~+~ \hat g_2 \, \cL_{\vec v}  \,\hat g_0 ~-~ \big(\vec \nabla_{\vec u}  \, \hat g_1\big)^2      ~=~ 0  \,.
\end{aligned}
 \label{gfunctioneqns}
\end{equation}
The first equation can be written 
\begin{equation}
\cL_{\vec v} \, \hat g_2 ~-~ \hat g_2^{3} \, \cL_{\vec u} (\hat g_2^{-1})     ~=~ 0 \,,
 \label{g2eqn-new}
\end{equation}
which leads to an obvious ``separable'' solution: 
\begin{equation}
\hat g_2      ~=~ \frac{h_2(\vec v)}{h_1(\vec u)} \,,
 \label{sep-sol}
\end{equation}
where the $h_j$ are harmonic.  This is the near-brane, limiting boundary condition discussed in   \cite{Lunin:2007mj}.  

With this choice for $\hat g_2$, the remaining equations in (\ref{gfunctioneqns}) are linear.  There is also the gauge redundancy associated with a linear shift $z \to z + const.$.  We will make the simple choice: $\hat g_1 \equiv 0$, which also eliminates the redundancy.  Hence we take 
\begin{equation}
G_0  ~=~  - \frac{1}{2} \, z^2 \,  \frac{h_2(\vec v)}{h_1(\vec u)}  ~+~ \hat g_0(u,v) \,, 
 \label{G0red}
\end{equation}
and the \mname\ equation, (\ref{master}), reduces to the requirement that the  $h_j$ are harmonic and that $\hat g_0$ satisfy the linear equation: 
\begin{equation}
\frac{1}{h_1(\vec  u) } \, \cL_{\vec u}  \,\hat g_0  ~+~  \frac{1}{h_2(\vec v)} \,    \cL_{\vec v} \, \hat g_0     ~=~ 0 \,,
 \label{g0eqn}
\end{equation}

Having got to this point we note that we now have:
\begin{equation}
w    ~=~ \partial_z G_0 ~=~  -  z  \,  \frac{h_2(\vec v)}{h_1(\vec u)} \,,
 \label{wform}
\end{equation}
and this means that the non-diagonal frame in the metric can be greatly simplified. Specifically: 
\begin{equation}
\begin{aligned}
 e^{-A_0} \, e^2 & ~=~  (-\partial_z w )^{\frac{1}{2}} \, \Big( dz ~+~(\partial_z w )^{-1}\,  \big (\vec \nabla_{\vec u} \, w \big)  \cdot  d \vec u \Big)  ~=~ \bigg( \frac{h_2(\vec v)}{h_1(\vec u)} \bigg)^{\frac{1}{2}} \bigg[ \, dz ~-~ \frac{z}{h_1(\vec u)}  \big (\vec \nabla_{\vec u} \, h_1(\vec u) \big)  \cdot  d \vec u \,  \bigg] \\
& ~=~ \big(h_1(\vec u) h_2(\vec v) \big)^{\frac{1}{2}} \Bigg[   \frac{dz }{h_1(\vec u)} ~-~ \frac{z}{(h_1(\vec u))^2}  \big (\vec \nabla_{\vec u} \, h_1(\vec u) \big)  \cdot  d \vec u   \Bigg]  ~=~ \big(h_1(\vec u) h_2(\vec v) \big)^{\frac{1}{2}}  \, d  \hat z    \,,
 \end{aligned}
 \label{fram2}
\end{equation}
where 
\begin{equation}
\hat z  ~\equiv~ \frac{z }{h_1(\vec u)}  \,. \label{zhatdefn}
\end{equation}
In other words, the fibration is ``pure gauge.'' Hence, both the fibration and the $z$-dependence of the metric can be removed by a judicious change of variable. It is the $\hat z$-coordinate that is the correct smeared M-theory direction.

There is probably a rich class of solutions to equation  (\ref{g0eqn}), but there is one interesting,  non-trivial way to satisfy it. One first re-writes   (\ref{g0eqn}) as:   
\begin{equation}
 \cL_{\vec u}  \,\hat g_0   ~=~  - h_0(\vec u, \vec v) \, h_1(\vec u) \,, \qquad   \cL_{\vec v}  \,\hat g_0   ~=~   h_0(\vec u, \vec v) \, h_2(\vec v)  \,,
 \label{g0solving}
\end{equation}
for some function, $h_0$.    One then follows \cite{deBoer:1999gea} by imposing the constraint $h_0 = h_1 h_2$ so that: 
\begin{equation}
\hat g_0  ~=~   f_2(\vec v) \, h_1(\vec u)~-~  f_1(\vec u) \, h_2(\vec v)  \,, \quad {\rm where} \quad   \cL_{\vec u}  \,f_1 ~=~h_1^2 \,, \quad  \cL_{\vec v}  \,f_2~=~h_2^2 \,, 
 \label{g0sol}
\end{equation}
and hence 
\begin{equation}
G_0   ~=~   - \frac{1}{2} \, z^2 \,  \frac{h_2(\vec v)}{h_1(\vec u)}  ~+~   f_2(\vec v) \, h_1(\vec u)~-~  f_1(\vec u) \, h_2(\vec v)  \,.
 \label{G0sol}
\end{equation}

Using this in  (\ref{solfns}), one obtains
\begin{equation}
w ~=~ -z \,  \frac{h_2(\vec v)}{h_1(\vec u)}  \,, \qquad  e^{-3  A_0} \, \bigg( \frac{h_2(\vec v)}{h_1(\vec u)} \bigg)^{\frac{1}{2}} ~=~  h_1(\vec u) \, h_2^2(\vec v) \quad \Rightarrow\quad
e^{- 2A_0} ~=~  h_1(\vec u) \, h_2(\vec v)  \,. \label{simpsolfns}
\end{equation}

The end result is precisely the family of solutions constructed in \cite{deBoer:1999gea} and, in particular, the metric reduces to:
\begin{equation}
ds_{11}^2 ~=~    (h_1(\vec u) \, h_2(\vec v))^{-1}\, (-dt^2 ~+~ dy^2 ) ~+~   d  \hat z^2 ~+~  h_1 (\vec u)  \, d \vec u \cdot d \vec u  ~+~ h_2\, (\vec v)\, d \vec v \cdot d \vec v \,.
 \label{11metric-simp}
\end{equation}

One should note that, if one starts from the more general framework of Section \ref{sec:GenRes}, then the independence from the M-theory direction and the removal of the non-trivial fibration requires a re-definition of the $z$-coordinate.
 
\subsection{More general families of solutions}
\label{ss:AdS3sols}

There are more general, ``unsmeared'' solutions that have been obtained in a ``near-brane'' limit \cite{Lunin:2007ab,DHoker:2008lup,DHoker:2008rje,DHoker:2008wvd,DHoker:2009lky, DHoker:2009wlx,Bachas:2013vza}.  Here we summarize the key results of \cite{Bachas:2013vza}.

The Ansatz makes full use of the isometries: 
\begin{equation}
\begin{aligned}
ds_{11}^2 &~=~   e^{2A} \, \big( \, \hat f_1^2 \, ds_{AdS_3}^2 ~+~ \hat f_2^2 \, ds_{S^3}^2 ~+~ \hat f_3^2 \, ds_{{S'}^3}^2 ~+~ h_{ij} d \sigma^i  d \sigma^j \,   \big)  \,, \\
C^{(3)}  &~=~  b_1\, \hat e^{012} ~+~ b_2\, \hat e^{345} ~+~ b_3\, \hat e^{678}  \,,
\end{aligned}
 \label{DHokerAnsatz}
\end{equation}
where the metrics $ds_{AdS_3}^2$, $s_{S^3}^2$  and $ds_{{S'}^3}^2$ are the metrics of unit radius on  AdS$_3$ and  the three-spheres and $\hat e^{012}$, $\hat e^{345}$ and $\hat e^{678}$ are the corresponding volume forms. 

The functions $e^{2A}$, $\hat f_j$, $b_j$, and the two-dimensional metric, $h_{ij}$, are, {\it a priori}, arbitrary functions of $(\sigma^1, \sigma^2)$ (and the $e^{2A}$ factor is redundant).  However, the final result in  \cite{Bachas:2013vza} is to pin down all these functions and express them in terms of a complex function, $G$, and a real function $h$.

First, the two dimensional metric must be that of a Riemann surface with K\"ahler potential, $\log(h)$:
\begin{equation}
h_{ij} d \sigma^i  d \sigma^j ~=~ \frac{\partial_w h \partial_{\bar w}   h }{  h^2}  \, |dw|^2    \,,
 \label{KahlerMet}
\end{equation}
where $w$ is a complex coordinate and  $h$ is required to be harmonic:
\begin{equation}
 \partial_w  \partial_{\bar w} h  ~=~ 0   \,.
 \label{harmonic}
\end{equation}
We will define real and imaginary parts of $w$ via:
\begin{equation}
w ~=~  \xi ~+~ i\, \rho \quad \Rightarrow \quad \partial_w ~=~  \coeff{1}{2}\, \big( \partial_\xi ~-~ i\, \partial_\rho \big) \,, \qquad \partial_{\bar{w}}~=~  \coeff{1}{2} \,\big( \partial_\xi ~+~ i\, \partial_\rho \big)  \,.
 \label{wparts}
\end{equation}
It is also convenient to introduce the harmonic conjugate, $\tilde h$, of $h$, defined by requiring that  $-\tilde h + i h$ is holomorphic:
\begin{equation}
 \partial_{\bar w}  (- \tilde h + i h)  ~=~   0 \,.
 \label{harmconj}
\end{equation}
Since $-\tilde h + i h$ is holomorphic we can use them as local coordinates on the Riemann surface, or, equivalently we can take 
\begin{equation}
- \tilde h ~+~ i h   ~=~   \beta \, w ~=~   \beta \, ( \xi ~+~ i\, \rho )\,,
 \label{simph}
\end{equation}
where $\beta$ is a constant parameter introduced for later convenience. 

Thus  we may (locally) fix  the Riemann surface metric to be a multiple of that of the Poincar\'e upper half-plane:  
\begin{equation}
h_{ij} d \sigma^i  d \sigma^j ~=~ \frac{ d\xi^2 ~+~ d\rho^2}{4 \, \rho^2}  \,,
 \label{PoincareMet}
\end{equation}
where the factor of $4$ comes from the factors of $\frac{1}{2}$ in partial derivatives (\ref{wparts}). 

The complex function, $G$, is required to satisfy the equation: 
\begin{equation}
 \partial_w   \, G ~=~  \coeff{1}{2}\, (\, G  +  \overline G\,) \,  \partial_w \log(h) \,. 
 \label{Geqn}
\end{equation}
If one writes $G$ in terms of real and imaginary parts, $G = g_1 + i g_2$, and uses the local coordinates (\ref{simph}), then one has:
\begin{equation}
 \partial_\xi  g_1 ~+~  \partial_\rho    g_2 ~=~ 0 \,,  \qquad  \partial_\xi  g_2 ~-~   \partial_\rho    g_1 ~=~ - \frac{1}{\rho} \, g_1\,. 
 \label{gjeqns1}
\end{equation}

It is convenient to introduce potentials, $\Phi$, and $\tilde \Phi$, associated with $G$.  First, one defines $\Phi$ via:
\begin{equation}
 \partial_w   \, \Phi  ~=~  \overline G \,  \partial_w  h  \qquad \Leftrightarrow \qquad \partial_\xi \Phi ~=~ - \beta \, g_2  \,, \quad  \partial_\rho \Phi ~=~ \beta \, g_1 \,. 
 \label{Phidefn}
\end{equation}
The existence of such a $\Phi$ is guaranteed by the first equation in (\ref{gjeqns1}). The second equation in (\ref{gjeqns1}) implies that $\Phi$ must satisfy
\begin{equation}
\Big(\partial_\xi^2  ~+~ \partial_\rho^2 ~-~ \frac{1}{\rho} \,  \partial_\rho \Big)   \, \Phi ~=~  0 \,. 
 \label{Phieqn}
\end{equation}

Similarly, the  second equation in (\ref{gjeqns1}) implies that there is a conjugate potential, $\tilde \Phi$, defined by:
\begin{equation}
\partial_\xi \tilde \Phi ~=~ - \frac{\beta}{\rho} \, g_1 ~=~ - \frac{1}{\rho} \, \partial_\rho \Phi     \,, \qquad  \partial_\rho \tilde \Phi ~=~  - \frac{\beta}{\rho} \, g_2 ~=~  \frac{1}{\rho} \, \partial_\xi \Phi  \,. 
 \label{Phidefns}
\end{equation}
The first equation in (\ref{gjeqns1}) then implies that $\tilde \Phi$ must satisfy
\begin{equation}
\partial_\xi^2 \tilde \Phi  ~+~ \frac{1}{\rho} \, \partial_\rho \big( \, \rho\,  \partial_\rho \tilde \Phi  \, \big)   \, ~=~  0 \,. 
 \label{tPhieqn}
\end{equation}

If one introduces a dummy coordinate, $\chi$, and considers Euclidean $\IR^3$ with coordinates $(\rho, \chi, \xi)$, where $\xi$ defines one of the axes and $(\rho, \chi)$ are polar coordinates in the remaining $\IR^2$, then the equation  (\ref{tPhieqn}) is simply the condition that $\tilde \Phi$ is harmonic on $\IR^3$.  Moreover, if one defines a one-form, $\Phi d \chi$, then  the relationship between $\Phi$ and $\tilde \Phi$  in (\ref{Phidefns}) can be summarized as $*_3 ( \Phi d \chi) = d \tilde \Phi$.  It remains to be seen if this is simply a coincidence or where there is some deeper physical meaning to this observation and the coordinate, $\chi$.
 
Finally, define the functions: 
\begin{equation}
W_\pm  ~\equiv~  | G ~\pm~ i |^2 ~+~ \gamma^{\pm 1} \, ( G    \overline G~-~1)   \,, 
 \label{Wpmdefn}
\end{equation}
where $-1 \le \gamma \le 1$ is a ``deformation'' parameter that defines the relevant exceptional superalgebra $D(2,1;\gamma)  \oplus D(2,1;\gamma)$ \cite{Bachas:2013vza}. 

The $\gamma =1$ limit will be essential to our subsequent analysis. Indeed, it was noted in \cite{DHoker:2008wvd,Bachas:2013vza} that the  superalgebra $D(2,1;\gamma)  \oplus D(2,1;\gamma)$ can only be embedded in $osp(8 |4, \IR)$  for $\gamma=1$, and in this limit the superalgebra becomes   $osp(4 |2, \IR) \oplus osp(4 |2, \IR)$.      Therefore, restricting to $\gamma =1$  is quite probably a crucial step if one is to embed the brane configuration into an asymptotically-flat background because the supersymmetries in asymptotically-flat geometries are  expected to be subalgebras of  $osp(8 |4, \IR)$.  

The sign of gamma is related to the magnitude of $G$ via:
\begin{equation}
 \gamma \, ( G    \overline G~-~1) ~\ge~ 0  \,,
 \label{gammacond}
\end{equation}
and to keep our presentation simple, we will henceforth restrict to:
\begin{equation}
 \gamma ~>~ 0\,, \qquad |G|  ~\ge~ 1  \,.
 \label{bounds1}
\end{equation}
With this choice, the parameters in \cite{Bachas:2013vza} can be simplified to:
\begin{equation}
c_1  =  \gamma^{1/ 2} + \gamma^{-1/ 2} > 0 \,,   \qquad c_2 =  -\gamma^{1/ 2} <  0  \,, \qquad c_3 = -\gamma^{-1/ 2}  < 0 \,, \qquad \sigma =+1  \,.
 \label{params}
\end{equation}

The metric functions in  (\ref{DHokerAnsatz})  are given by: 
\begin{equation}
 \hat f_1^{-2}   ~=~   \gamma^{-1} \, (\gamma +1)^2 \,  ( G    \overline G~-~1)  \,, \qquad  \hat f_2^{-2}    ~=~  W_+ \,, \qquad \hat f_3^{-2}   ~=~  W_-  \,,  \ 
 \label{f-functions}
\end{equation}
and
\begin{equation}
\begin{aligned}
  e^{6A}   ~=~  h^2\, ( G    \overline G~-~1)  \, W_+\, W_-   ~=~ \gamma \, (\gamma +1)^{-2}  \,   h^2\,  \hat f_1^{-2}   \,  \hat f_2^{-2}  \,  \hat f_3^{-2}      \,.
\end{aligned}
 \label{eA-function}
\end{equation}

The flux functions, $b_i$, are given by:
\begin{equation}
\begin{aligned}
b _1  &~=~  \frac{\nu _1}{c_1^3}\, \bigg[\,    \frac{ h \, (G + \overline G) }{(G    \overline G~-~1) } ~+~\gamma^{-1} \, (\gamma +1)^2 \, \Phi ~-~ (\gamma - \gamma^{-1})\, \tilde h  \,\bigg]\,, \\
b _2  &~=~  \frac{  \nu _2}{c_2^3}\, \bigg[\,   -\frac{ h \, (G + \overline G) }{W_+ } ~+~(\Phi ~-~\tilde h ) \,\bigg]\,,  \qquad 
b _3 ~=~   \frac{\nu _3}{c_3^3}\, \bigg[\,    \frac{ h \, (G + \overline G) }{  W_-} ~-~ (\Phi ~+~\tilde h )\,\bigg]\,, 
\end{aligned}
\label{bfunctions}
\end{equation}
where one has $|\nu_i| =1$, with signs arranged so that $\nu_1\nu_2\nu_3= - 1$.  Compared to the results in  \cite{Bachas:2013vza}, we have used (\ref{params}) and
 we have dropped some (irrelevant) additive constants in the $b_i$.

\subsection{Mapping the \texorpdfstring{AdS$_3$}{AdS3} solutions to M5-M2 intersections}
\label{ss:mapping}

Our goal here is to show how to map the AdS$_3$ solutions of Section \ref{ss:AdS3sols} into the spherically-symmetric brane-intersection formulation of Section \ref{ss:sphsymm}.  

The first step is to remember that the AdS$_3$ solutions, supposed to correspond to a near-brane limit, depend non-trivially on only two coordinates, $(\rho, \xi)$, whereas the formulation in Section  \ref{ss:sphsymm} allows asymptotically-flat solutions that depend non-trivially on three variables, $(u,v,z)$.  We thus have to find a scaling limit for the solutions in Section \ref{ss:sphsymm}.

To render the scaling properties more transparent, we introduce a Poincar\'e metric on AdS$_3$:
\begin{equation}
 ds_{AdS_3}^2  ~=~   \frac{d \mu^2}{\mu^2} ~+~ \mu^2 \, \big(-dt^2 ~+~ dy^2 \big)\,,
 \label{PoinAdS}
\end{equation}
where the Poincar\'e $\IR^{1,1}$ factor represents the common directions of the brane intersection, and it is to be identified with the same factor in  (\ref{11metric-symm}).

We start by noting that the metric  (\ref{DHokerAnsatz}) is scale invariant under:
\begin{equation}
\mu   ~\to~  \lambda \,\mu  \,, \qquad (t,y) ~\to~  \lambda^{-1} (t,y) \,.
 \label{scaling1}
\end{equation}
This must now be imposed on the more general class of solutions discussed in Section \ref{ss:sphsymm}.

Scale invariance of  (\ref{11metric-symm}) can be achieved by taking: 
\begin{equation}
(u ,v)  ~\to~ \sqrt{ \lambda} \,(u ,v)   \,, \qquad z   ~\to~  \lambda^{-1} z  \,,
 \label{scaling2}
\end{equation}
and 
\begin{equation}
e^{A_0} ~\to~ \lambda\, e^{A_0}  \,, \qquad  w   ~\to~  \lambda^{-1} \, w  \,.
 \label{scaling3}
\end{equation}

There is a very important difference between  (\ref{scaling1}),  (\ref{scaling2}) and  (\ref{scaling3}).  The first two are simply prescriptions for the scaling of coordinates, while the (\ref{scaling3}) imposes strong constraints  on the functional form of $e^{A_0}$ and $w$.  It is these constraints that lead to the near-brane limit.  Indeed, this scaling invariance leads to the following Ansatz for the mapping we seek:
\begin{equation}
\begin{aligned}
u  &~=~ \sqrt {\mu } \, m_1(\rho, \xi)  \,,   \qquad  v  ~=~ \sqrt {\mu } \, m_2 (\rho, \xi)\,,   \qquad  z  ~=~\mu^{-1} \, m_3 (\rho, \xi) \,,   \\
  w  &~=~\mu^{-1} \, m_4 (\rho, \xi) \,,   \qquad  e^{A_0}  ~=~\mu  \, m_5 (\rho, \xi)   \,,
\end{aligned}
 \label{variables1}
\end{equation}
for some functions, $m_j$.

A direct comparison of  (\ref{11metric-symm}) and  (\ref{DHokerAnsatz}), using (\ref{PoinAdS}) and  (\ref{PoincareMet}), leads to:
\begin{equation}
 e^{2A}\, \hat f_1^2\, \mu^2  ~=~ e^{2A_0}  \,, \qquad   e^{2A}\, \hat f_2^2   ~=~ e^{- A_0}  \,(-\partial_z w)^{-\frac{1}{2}} \,  u^2  \,, \qquad   e^{2A}\, \hat f_3^2   ~=~ e^{- A_0}  \,(-\partial_z w)^{\frac{1}{2}} \,  v^2   \,,
 \label{identities1}
\end{equation}
along with 
\begin{equation}
\begin{aligned}
e^{2A} \, \bigg( \, \hat f_1^2 \,  \frac{d \mu^2}{\mu^2}  ~+~ \frac{ d\xi^2 ~+~ d\rho^2}{4 \, \rho^2}   \bigg) ~=~  & e^{- A_0}  \Big( (-\partial_z w)^{-\frac{1}{2}} \, du^2~+~  (-\partial_z w)^{\frac{1}{2}} \, dv^2  \Big)  \\
& +~  e^{2 A_0} \,  (-\partial_z w ) \, \Big( dz ~+~(-\partial_z w )^{-1}\,  \big (\partial_u w \big)  \, d u \Big)^2 \,.
\end{aligned}
 \label{identities2}
\end{equation}
Using   (\ref{simph}), (\ref{f-functions}),  (\ref{eA-function}) and  (\ref{identities1}) in (\ref{identities2}) one finds that one must have:
\begin{equation}
\begin{aligned}
   \frac{\gamma}{( 1+\gamma)^2 }  & \,    \frac{1}{( G    \overline G -1) }   \,  \frac{d \mu^2}{\mu^2}  ~+~ \frac{ d\xi^2 ~+~ d\rho^2}{4 \, \rho^2} \\
~=~  &   \frac{1}{ W_+}   \, \frac{du^2}{u^2} ~+~    \frac{1}{ W_-}  \,  \frac{dv^2}{v^2}   ~+~ \frac{1}{\beta^2\,  \rho^2\,  ( G    \overline G -1) }  \frac{W_+}{ W_- }   \, \bigg( u^2 dz ~+~(\partial_z w )^{-1}\,  \big (u^3 \partial_u w \big)  \, \frac{d u}{u} \bigg)^2 \,.
\end{aligned}
 \label{identities3}
\end{equation}
One can also manipulate  (\ref{identities1}), using (\ref{eA-function}) and (\ref{simph}),  to obtain:
\begin{equation}
u^2 v^2  ~=~ \frac{\beta^2 \, \gamma}{(\gamma +1)^2} \,\,\mu^2  \, \rho^2    \,, \qquad    (-\partial_z w)\, \frac{v^2}{u^2} ~=~ \frac{W_+}{ W_-}   \,,\qquad   e^{A_0}~=~  \frac{ \beta  \sqrt{\gamma} \mu \rho}{(\gamma +1)} e^{-2 A}  (W_+ W_-)^{\frac{1}{2}} \,.
\label{identities4}
\end{equation}

The first identity in  (\ref{identities4}),  and the form of the fibration on the right-hand side of  (\ref{identities3}), suggest a slightly more refined change of variables:
\begin{equation}
u  ~=~  \sqrt{a\, \mu \, \rho}  \, e^{\alpha (\rho, \xi)}  \,,   \qquad  v  ~=~  \sqrt{a\, \mu \, \rho}    \, e^{-\alpha (\rho, \xi)}   \,,   \qquad  z  ~=~\mu^{-1} \,e^{-2\alpha (\rho, \xi)}  \,  p (\rho, \xi) \,,
 \label{variables2}
\end{equation}
where $\alpha$ and $p$ are  functions to be determined,  and the parameter $a$ is given by:
\begin{equation}
a~\equiv~  \frac{\beta \, \sqrt{\gamma}}{(\gamma +1)} \,.
 \end{equation}

Comparing the expressions (\ref{C3symm}) and (\ref{DHokerAnsatz}) for the $C^{(3)}$ flux leads to:
\begin{equation}
(\partial_z w )^{-1} \, \big (u^3 \partial_u w \big) ~=~ b_2   \,, \qquad    \big(v^3 \partial_v w \big) ~=~ b_3  \,.
\label{identities5}
 \end{equation}
where $b_2$ and $b_3$ are given in (\ref{bfunctions}).  The matching of the flux along the AdS direction involves non-trivial gauge transformations and we will return to this below. 

From this we note that  (\ref{identities3}) can be re-written as
\begin{equation}
\begin{aligned}
   \frac{\gamma}{( 1+\gamma)^2 }  & \,    \frac{1}{( G    \overline G -1) }   \,  \frac{d \mu^2}{\mu^2}  ~+~ \frac{ d\xi^2 ~+~ d\rho^2}{4 \, \rho^2} \\
~=~  &   \frac{1}{ W_+}   \, \frac{du^2}{u^2} ~+~    \frac{1}{ W_-}  \,  \frac{dv^2}{v^2}   ~+~ \frac{1}{\beta^2\,  \rho^2\,  ( G    \overline G -1) }  \frac{W_+}{ W_- }   \, \bigg( u^2 dz ~+~ b_2 \, \frac{d u}{u} \bigg)^2 \,.
\end{aligned}
 \label{identities6}
\end{equation}
Substituting the change of variable (\ref{variables2}) into this, one obtains an over-determined system of equations for $b_2$ and the derivatives of $\alpha$ and $p$.
This system is very complicated, involving square-roots of a quadratic in $W_\pm$. However, for $\gamma=1$, the system dramatically simplifies and one finds:
\begin{equation}
\begin{aligned}
  \partial_\xi \alpha &~=~ -\frac{\varepsilon_1}{2 \rho} \, g_1 \,,  \quad \partial_\rho \alpha ~=~ \frac{1}{2 \rho} \, g_2 \,, \qquad b_2 ~=~ 2 \, a \,\rho \, p ~+~   \frac{\varepsilon_2 \,\beta \rho\, g_1}{g_1^2 + g_2^2 + g_2} \,,   \\
   \partial_\xi p &~=~ -\frac{\varepsilon_1 \varepsilon_2 \,\beta}{2\, a \rho} \, (g_2 -1) \,,  \quad \partial_\rho p ~=~ -\frac{1}{\rho} \,\bigg(  p~+~ \frac{\varepsilon_2 \,\beta}{2\,a} \, g_1   \bigg)\,.
\end{aligned}
 \label{diffconstraints}
\end{equation}
where $g_1$ and  $g_2$  are the real and imaginary parts of $G$, $G = g_1 + i g_2$.

From  (\ref{gjeqns1}) one sees that $\partial_\xi (\rho^{-1} g_2) = \partial_\rho (\rho^{-1} g_1)$ and hence we must take $\varepsilon_1 =- 1$, and then one can identify $\alpha$ with the potential $\tilde \Phi$:
\begin{equation}
\alpha~=~ - \frac{1}{2\, \beta} \,  \tilde \Phi \,.
\label{alphares}
 \end{equation}
Similarly, it is elementary to integrate the equations for $p$ to arrive at:
\begin{equation}
p~=~ -  \frac{\varepsilon_2}{2\, a\,\rho} \,  \big(  \Phi ~+~ \beta \, \xi \big)\,.
\label{pres}
 \end{equation}
Using  (\ref{diffconstraints}) one finds that $b_2$ must have the form:
\begin{equation}
b_2 ~=~  \varepsilon_2 \, \bigg( \frac{ \beta \rho\, g_1}{g_1^2 + g_2^2 + g_2} ~-~ \big(  \Phi + \beta \, \xi \big)  \bigg)~=~   \varepsilon_2 \, \bigg( \frac{ h \,(G + \overline G)}{W_+} ~-~ \big(  \Phi - \tilde h \big)  \bigg) \,.
\label{b2deduced}
 \end{equation}
From  (\ref{params}) one sees that $c_2 =-1$ for $\gamma =1$, and one finds a perfect match between (\ref{b2deduced}) and (\ref{bfunctions}) if $\nu_2 =  \varepsilon_2$.

To summarize, in order to map the solution in Section \ref{ss:AdS3sols} to the near-brane limit of the spherically-symmetric brane-intersection of Section \ref{ss:sphsymm} one needs to take:
\begin{equation}
\gamma = 1 \,, \qquad u  = \big(\coeff{1}{2}\, \beta \mu  \rho\big)^{\frac{1}{2}}  \, e^{- \frac{1}{2\, \beta} \,  \tilde \Phi}  \,,   \qquad  v  =  \big(\coeff{1}{2}\, \beta \mu  \rho\big)^{\frac{1}{2}}   \,  e^{+ \frac{1}{2\, \beta} \,  \tilde \Phi}   \,,   \qquad  z  =  - \frac{\varepsilon_2}{\beta \rho \mu} \,   e^{\frac{1}{\beta} \,  \tilde \Phi }  \,    \big(  \Phi +  \beta \, \xi \big) \,.
 \label{variables3}
\end{equation}

One can also compute $w$ as a function of $(\mu, \rho, \xi)$.  Indeed, from  (\ref{identities4}) and (\ref{identities5}) one has
\begin{equation}
\partial_z w ~=~ - \frac{g_1^2 + g_2^2+ g_2}{ g_1^2 + g_2^2 - g_2} \, e^{- \frac{2}{\beta}\,  \tilde \Phi}    \,, \qquad(\partial_z w )^{-1} \, \big (u^3 \partial_u w \big) ~=~ b_2   \,, \qquad    \big(v^3 \partial_v w \big) ~=~ b_3  \,,
\label{widentities}
 \end{equation}
from which one obtains:
\begin{equation}
dw ~=~ (\partial_z w) dz  ~+~  ( \partial_u w) du ~+~  ( \partial_v w) dv ~=~ d\,\bigg[ \,  \frac{\varepsilon_2}{\beta \rho \mu}  \, e^{- \frac{1}{ \beta} \,  \tilde \Phi}  (\Phi - \beta \xi) \, \bigg]   \,,
\label{dwexpression}
 \end{equation}
and hence:
\begin{equation}
w ~=~\frac{\varepsilon_2}{\beta \rho \mu}  \, e^{- \frac{1}{ \beta} \,  \tilde \Phi}  (\Phi- \beta \xi) \,.
\label{wres}
 \end{equation}

To get an exact differential on the right-hand side of (\ref{dwexpression})  it is essential that one has 
\begin{equation}
b_3  ~=~  \varepsilon_2 \, \bigg( \frac{ \beta \rho\, g_1}{g_1^2 + g_2^2 - g_2} ~-~ \big(  \Phi - \beta \, \xi \big)  \bigg)~=~  \varepsilon_2 \, \bigg( \frac{ h \,(G + \overline G)}{W_-} ~-~ \big(  \Phi +  \tilde h \big)  \bigg) \,.
\label{b3deduced}
 \end{equation}
From  (\ref{params}) one sees that $c_3 =-1$ for $\gamma =1$, and one finds a perfect match between (\ref{b3deduced}) and (\ref{bfunctions}) if $\nu_3 =  -\varepsilon_2$.

It is interesting to note that (\ref{variables3}) and (\ref{wres}) imply
\begin{equation}
 u^2 \, z  ~=~  - \coeff{1}{2}\,   \varepsilon_2  \,    \big(  \Phi + \beta \, \xi \big)  \,, \qquad v^2 \, w  ~=~ \coeff{1}{2}\,   \varepsilon_2  \,    \big(  \Phi - \beta \, \xi \big) \,,
 \label{zwsimp}
\end{equation}
which, once again, illustrates the ``democracy'' in the fibration discussed in Appendix \ref{app:democracy}.  Additionally, it is  interesting to observe that   (\ref{variables3}) and  (\ref{wres})  imply that if one flips the signs of the potentials,  then one flips the roles of $u$ and $v$ and the roles of $z$ and $w$: 
\begin{equation}
 \Phi \to -\Phi\,, \  \ \tilde \Phi  \to -\tilde \Phi  \quad \Rightarrow \quad  u  \leftrightarrow v \,, \quad    z  \leftrightarrow  w   \,.
 \label{interchanges}
\end{equation}

Finally, consider the differential: 
\begin{equation}
\omega ~\equiv~ e^{3  A_0}\, (-\partial_z w )^{\frac{1}{2}} \, \Big( dz ~+~(\partial_z w )^{-1}\,  \big (\partial_u w \big)  \, d u \Big)   \,.
 \label{differential1}
\end{equation}
Using (\ref{eA-function}), (\ref{identities4})  and (\ref{variables3}) one finds:  
\begin{equation}
\begin{aligned}
\omega & ~=~  \frac{W_+ \, \mu^2  }{4 \, (G \overline G-1)}\,   \bigg( u^2 dz ~+~(\partial_z w )^{-1}\,  \big (u^3 \partial_u w \big)  \, \frac{d u}{u} \bigg) \\
 & ~=~   \frac{\varepsilon_2 }{4}\, \bigg[ \bigg( \frac{ \beta \rho\, g_1}{g_1^2 + g_2^2 - 1} ~+~  2\, \Phi \bigg) \, \mu \, d \mu ~-~  d\, \big( \mu^2\, \Phi \big) \bigg]\\
  & ~=~   \frac{\varepsilon_2}{8}\,\bigg[ \bigg( \frac{h  (G + \overline G)}{(G \overline G-1)} ~+~  4\, \Phi \bigg)  \, \mu \, d \mu~-~  d\, \big( 2\, \mu^2\, \Phi \big)\bigg] ~=~  \frac{\varepsilon_2}{\nu_1}\, b_1 \, \mu \, d \mu ~-~  \frac{\varepsilon_2}{4}\, d\, \big( \mu^2\, \Phi \big)  \,,
\end{aligned}
 \label{omegaform}
\end{equation}
where the last expression follows from (\ref{bfunctions})  and  (\ref{params}) for $\gamma =1$.  Thus one has 
\begin{equation}
b_1  ~=~  \frac{\nu_1}{4}  \, \bigg( \frac{ \beta \rho\, g_1}{g_1^2 + g_2^2 - 1} ~+~  2\, \Phi    \bigg) \,,
\label{b1deduced}
 \end{equation}
as in (\ref{bfunctions}) and \cite{Bachas:2013vza} with $c_1 =2$.

The important point here is that the three-form potential  (\ref{C3gen})  along $y$ (the common M2-M5 direction) and $z$ (the M-theory direction) is:
\begin{equation}
C_{tyz}^{(3)} ~=~   - e^0 \wedge e^1 \wedge e^2  ~=~  -dt \wedge dy \wedge \omega  ~=~ - \frac{\varepsilon_2}{\nu_1}\, b_1 \, \mu \,  dt \wedge dy \wedge d \mu ~+~  \frac{\varepsilon_2}{4}\, d\, \big( \mu^2\, \Phi  \, dt   \wedge dy  \big) \,.
 \label{C3symmtyz}
\end{equation}
Using (\ref{DHokerAnsatz}) and  (\ref{PoinAdS})  one has:
\begin{equation}
C_{tyz}^{(3)} ~=~   b_1 \, \mu \,  dt \wedge dy \wedge d \mu \,,
 \label{C3tyz-DHoker}
\end{equation}
and so these components of the flux match (\ref{C3symmtyz}), up to a gauge transformation, provided that $\nu_1 = - \varepsilon_2$.

We have shown that there is perfect agreement if and only if $-\nu_1 = \nu_2 = -\nu_3 =  \varepsilon_2$.

 In  \cite{Bachas:2013vza} these is a further constraint  $\nu_1 \nu_2 \nu_3 = -1$, which  suggests $\varepsilon_2 = -1$, however this last constraint is related to the form of the  unbroken supersymmetries.  We will discuss all these signs in the next section.

\subsection{Verifying the BPS equations for the \texorpdfstring{AdS$_3$}{AdS3} solutions}
\label{ss:AdS-BPS}

We can use the results of Section (\ref{ss:mapping}) to verify that the AdS$_3$ solutions satisfy directly the BPS equations in Section  (\ref{sec:GenRes}).  Specifically, we have taken the following frames:
\begin{equation}
\begin{aligned}
e^0 ~=~  &\frac{\mu \, e^{A}}{2\,\sqrt{G \bar G-1}} \, dt \,, \qquad e^1~=~ \frac{\mu \, e^{A}}{2\,\sqrt{G \bar G-1}}\,dy \,, \\
e^2 ~=~  & \frac{ \varepsilon_2  \, e^{A}}{\rho \,  \sqrt{(G \bar G-1)\,W_+ \,W_-  }} \, \bigg( \rho\, g_1\, \frac{d \mu}{\mu} +  (G \bar G-1) \big(g_2 \, d\xi - g_1 \, d\rho \big) \bigg) \,, \\
e^3~=~  & \frac{ e^{A}}{2 \,  \sqrt{W_+}} \, \bigg( \frac{d \mu}{\mu}  + \frac{d \rho}{\rho}  +   \frac{1}{\rho} \big(g_1 \, d\xi  + g_2 \, d\rho \big) \bigg) \,, \\
 e^4~=~  & \frac{ e^{A}}{2 \,  \sqrt{W_-}} \, \bigg( \frac{d \mu}{\mu}  + \frac{d \rho}{\rho}  -   \frac{1}{\rho} \big(g_1 \, d\xi  + g_2 \, d\rho \big) \bigg) \,, \\
e^{i+4}~=~  &  \frac{ e^{A}}{2 \,  \sqrt{W_+}} \,    \sigma_i  \,, \qquad e^{i+7} ~=~    \frac{ e^{A}}{2 \,  \sqrt{W_-}} \, \tilde   \sigma_i   \,,   \qquad {i = 1,2,3}   \,.
\end{aligned}
 \label{11frames2}
\end{equation}
These are, in fact, precisely the same frames as in  (\ref{firstframes}); note, in particular, the possible sign choice, $\varepsilon_2$, in $e^2$.   Using (\ref{DHokerAnsatz}) to define the Ansatz for the flux, $C^{(3)}$, one finds that all the BPS equations can be satisfied if:
\begin{equation}
\begin{aligned}
\partial_\xi b_1  ~=~  &\varepsilon_2  \,  \partial_\xi \bigg[ - \frac{\beta  \, \rho\, g_1}{4 \,(G \bar G-1)}  \bigg]  ~+~    \frac{1}{2} \, \varepsilon_2  \, \beta \, g_2\,, \qquad   \partial_\rho b_1  ~=~\varepsilon_2  \,  \partial_\rho \bigg[ - \frac{\beta  \, \rho\, g_1}{4 \,(G \bar G-1)}  \bigg]  ~-~    \frac{1}{2} \, \varepsilon_2  \, \beta \, g_1  \,, \\
\partial_\xi b_2  ~=~  &\varepsilon_2  \,  \partial_\xi \bigg[ - \frac{2\, \beta  \, \rho\, g_1}{W_+}  +    \beta \,  \xi  \bigg]  ~-~    \varepsilon_2  \, \beta \, g_2\,, \qquad \partial_\rho b_2 ~=~\varepsilon_2  \, \partial_\rho \bigg[ - \frac{2\, \beta  \, \rho\, g_1}{W_+}  +     \varepsilon_2  \,  \beta \,  \xi  \bigg]  ~+~  \varepsilon_2  \,   \beta \, g_1  \,, \\
\partial_\xi b_3 ~=~  &\varepsilon_2  \, \partial_\xi \bigg[ - \frac{2\, \beta  \, \rho\, g_1}{W_-} -  \beta \,  \xi  \bigg]  ~-~  \varepsilon_2  \,   \beta \, g_2\,, \qquad \partial_\rho b_3~=~\varepsilon_2  \,  \partial_\rho \bigg[ - \frac{2\, \beta  \, \rho\, g_1}{W_-}  -   \beta \,  \xi  \bigg]  ~+~  \varepsilon_2  \,  \beta \, g_1  \,.
\end{aligned}
 \label{bderivs}
\end{equation}
which leads to:
\begin{equation}
b_1  =  - \varepsilon_2  \, \bigg(\frac{\beta  \, \rho\, g_1}{4 \,(G \bar G-1)}   +   \frac{1}{2} \,  \Phi \bigg) \,, \ \ 
 b_2  = - \varepsilon_2  \, \bigg( \frac{2\, \beta  \, \rho\, g_1}{W_+}   -    (\Phi + \beta \,  \xi ) \bigg) \,, \ \ 
b_3 =  - \varepsilon_2  \, \bigg(  \frac{2\,  \beta  \, \rho\, g_1}{W_-} -  (\Phi - \beta \,  \xi )\bigg)   \,.
 \label{bsols1}
\end{equation}
This is consistent with (\ref{bfunctions}) for $c_1=2, c_2 =c_3 =-1$,  if one takes $-\nu_1 = -\nu_2 = \nu_3 = \varepsilon_2$.

The signs of the fluxes determine the unbroken supersymmetries.  Indeed, if we generalize  (\ref{projs3}) to
\begin{equation}
 \Gamma^{012} \, \varepsilon  ~=~  \eta_1 \, \varepsilon \,,   \qquad  \Gamma^{013567} \, \varepsilon  ~=~  \eta_2 \, \varepsilon \,,   \qquad  \Gamma^{01489\,10} \, \varepsilon  ~=~-  \eta_1\, \eta_2 \,\varepsilon  \,.
 \label{projs4}
\end{equation}
where $ \eta_j $ are signs, $ \eta_j^2  =1$, then one finds:
\begin{equation}
\begin{aligned}
b_1  &~=~  \varepsilon_2  \, \eta_1 \bigg( \frac{\beta  \, \rho\, g_1}{4 \,(G \bar G-1)}   +    \frac{1}{2} \,  \Phi \bigg) \,, \qquad 
 b_2  ~=~   \varepsilon_2  \,  \eta_1  \eta_2 \,\bigg(\frac{2\, \beta  \, \rho\, g_1}{W_+}   -   (\Phi + \beta \,  \xi ) \bigg) \,, \\ 
b_3 &~=~  - \varepsilon_2  \,  \eta_2 \bigg(\frac{2\,  \beta  \, \rho\, g_1}{W_-} -  (\Phi - \beta \,  \xi ) \bigg)    \,.
\end{aligned}
 \label{bsols2}
\end{equation}
This matches (\ref{bfunctions}) for $c_1=2, c_2 =c_3 =-1$, if $ \nu_1=\varepsilon_2 \eta_1$, $ \nu_2  = \varepsilon_2 \eta_1  \eta_2$ and $ \nu_3 =\varepsilon_2 \eta_2 $.  Note that this implies  $\nu_1 \nu_2 \nu_3 = \varepsilon_2 $, and, as noted above, this matches the constraint in  \cite{Bachas:2013vza} if $\varepsilon_2 =-1$.  This means that the frame, $e^2$, in (\ref{11frames2}) comes with a negative sign.   Indeed, in Section \ref{ss:mapping} the matching required that $-\nu_1 = \nu_2 =- \nu_3 = \varepsilon_2$, which corresponds to  $\eta_1 = \eta_2  = \varepsilon_2 =-1$.

The choices of signs $\eta_1,  \eta_2$ and $\varepsilon_2$  are determined by the unbroken supersymmetry and frame orientations.   We will persist with the choice that we started with in (\ref{projs1}) and we will keep our frames positive.  Thus we will take:
\begin{equation}
\eta_1 ~=~-1\,, \qquad \eta_2 ~=~ +1\,, \qquad \varepsilon_2~=~+1\  \,.
 \label{signchoices}
\end{equation}
%

\section{String webs}
\label{sec:stringwebs}

Some of the solutions we have constructed can  be related, via dualities, to the $(p,q)$ string-web solutions preserving eight supercharges  that were obtained in \cite{Lunin:2008tf}. This relation is expected: when the M2-M5 solutions are smeared over three of the internal torus directions, one can find a duality chain that relates the M2 and M5 branes to F1 and D1 strings. To illustrate this connection and some of the interesting properties of our solutions it reveals, we perform the explicit duality at the level of supergravity solutions, and relate the F1-D1 string-web solutions of \cite{Lunin:2008tf} to the ones we obtained in Section \ref{sec:GenRes}. 

We begin with the string-web solutions: 
\begin{align}
\label{Lunin}
    &ds_{IIB}^2=\sqrt{h_{11}}\left[ -e^{3A}dt^2+e^{3A}h_{ab}dr^adr^b+\frac{e^{-3A}}{\det h}dw_2^2+d\boldsymbol{y}_6^2 \right] \,, \\
    e^{2\phi}&=\frac{h_{11}^2}{\det h}\,, \hspace{10pt} C_0=-\frac{h_{12}}{h_{11}}\,,  \hspace{10pt} B_2=e^{3A}h_{1a}dt\wedge dr^a,   \hspace{10pt}C_2=e^{3A}h_{2a}dt\wedge dr^a \nonumber
\end{align}
which describe a web of F1-strings, D1-strings and more generic $(p,q)$ strings in the plane spanned by $(r^1,r^2)$, with $w_2$ and $\textbf{y}_6$ being orthogonal directions. The two-dimensional metric $h_{ab}$ can be expressed in terms of a K\"ahler potential:
\begin{equation}
\label{Kahler}
	h_{ab}=\frac{1}{2}\partial_a \partial_b K(r^1,r^2,\boldsymbol{y}) \,,
\end{equation}
which satisfies a Monge-Amp\`ere equation
\begin{equation}
\label{MongeAmpere}
	\Delta_y K + 2 e^{-3A}=0 \,,
\end{equation}
with $e^{-3A}=\det h$. 	

\subsection{The D2-D4 frame}
\label{subsec:D2D4}

In order to dualize the solution \eqref{Lunin} to the M2-M5 duality frame and compare it with our solution \eqref{11metric}-\eqref{C3gen}, we have to first go to the D2-D4 frame by performing T-dualities along $w_2$ and $y_1$, an S-duality, and finally another T-duality along $y_2$.  Anticipating the mapping of this solution to the one of Section \ref{sec:GenRes}, we do the following coordinate relabeling: 
\begin{equation}
	r^1 \rightarrow z\,, \quad r^2 \rightarrow u_1\,, \quad w_2 \rightarrow u_2\,, \quad y_1 \rightarrow u_3\,, \quad y_2 \rightarrow y \,, \quad y_{3,4,5,6} \rightarrow \upsilon_{3,4,5,6} \,.
\label{relabelling}
\end{equation}

The procedure, explained in detail in Appendix \ref{app:duality-D2-D4}, involves working in the democratic formalism  and using several times the Monge-Amp\`ere equation \eqref{MongeAmpere}. The final result for the solution describing a D2-D4 web is
\begin{align}
\label{D2-D4 final}
    ds^2&=\frac{1}{\sqrt{\det h}}(-dt^2+dy^2) + \frac{\sqrt{\det h}}{h_{11}}(du_2^2+du_3^2) +\sqrt{\det h}\left(e^{3A}h_{ab}dr^adr^b+ds^2_{\mathbb{R}^4} \right)\,, \nonumber \\
    e^{2\phi}&=\frac{\sqrt{\det h}}{h_{11}}\,, \hspace{10pt} B_2=\frac{h_{12}}{h_{11}}du_2 \wedge du_3\,, \\
    C_3&=e^{3A}h_{1a}dt\wedge dr^a \wedge dy- \frac{\upsilon^3}{2}\partial_{\upsilon} \partial_z K \, d\Omega_3' \,, \nonumber \\ 
    C_5&=\frac{1}{h_{11}}dt\wedge du_1 \wedge du_2 \wedge du_3 \wedge dy + \frac{\upsilon^3}{2}\left( \frac{h_{12}}{h_{11}}\partial_{\upsilon} \partial_z K- \partial_{\upsilon} \partial_{u_1}K \right) du_2 \wedge du_3 \wedge d\Omega_3'\,. \nonumber
\end{align}

\subsection{Uplifting to M-theory}
In order to go to M-theory we have to uplift the system \eqref{D2-D4 final} along a direction, $x$, that we will call $x=-u_4$. Using the usual relations between type IIA and 11-dimensional supergravity \ 
\begin{align}
\label{IIA-M}
	ds_{11}^2&=e^{\frac{-2\phi}{3}}ds_{10}^2+e^{\frac{4\phi}{3}}(dx+C_1)^2 \,, \\
	C_3'&=C_3+B_2 \wedge dx \,,
\end{align}
where $x$ is the uplifting direction, we arrive at the following M-theory solution: 
\begin{align}
\label{M sol1}
	ds_{11}^2=&\frac{h_{11}^{1/3}}{(\det h)^{2/3}}(-dt^2+dy^2)+\frac{(\det h)^{1/3}}{h_{11}^{2/3}}(du_2^2+du_3^2+du_4^2) \nonumber \\ 
	&+(\det h)^{1/3}h_{11}^{1/3}\left(e^{3A}h_{ab}dr^a dr^b+ds^2_{\mathbb{R}^4} \right) \,, \\
\label{M sol2}
	C_3=&\frac{h_{11}}{\det h}dt\wedge dz \wedge dy +\frac{h_{12}}{\det h}dt\wedge du_1 \wedge dy -\frac{h_{12}}{h_{11}}du_2 \wedge du_3 \wedge du_4 -\frac{\upsilon^3}{2}\partial_{\upsilon} \partial_z K d\Omega_3'\,.
\end{align}
Remembering the re-labelling of the $r^a$ coordinates in (\ref{relabelling}), and adding and subtracting $\frac{(\det h)^{1/3}}{h_{11}^{2/3}}du_1^2$ in \eqref{M sol1}, the foregoing metric becomes:
\begin{align}
\label{M met}
	ds_{11}^2=&\frac{h_{11}^{1/3}}{(\det h)^{2/3}}(-dt^2+dy^2)+\frac{(\det h)^{1/3}}{h_{11}^{2/3}}(du_1^2+du_2^2+du_3^2+du_4^2) \nonumber \\ 
	&+\frac{h_{11}^{4/3}}{(\det h)^{2/3}}\left( dz+\frac{h_{12}}{h_{11}}du_1 \right)^2 +(\det h)^{1/3}h_{11}^{1/3}\left(d\upsilon^2+\upsilon^2 d\Omega_3'^2\right) \,,
\end{align}
where we have written the $\mathbb{R}^4$ metric of \eqref{M sol1} in hyperspherical coordinates, 

Now, we would like to compare this solution with \eqref{11metric}-\eqref{C3gen}. In order to do that we have to assume spherical symmetry in $\mathbb{R}^4(\vec{\upsilon})$\footnote{It is not necessary to assume spherical symmetry in $\mathbb{R}^4(\vec{\upsilon})$ in \ref{subsec:D2D4}, but doing so simplifies considerably the transition to the democratic formalism.} and fiber the ``original" M-theory direction, $z$, over a single direction of the $\mathbb{R}^4(\vec{u})$, which we pick, for obvious reasons, to be $u_1$. The metric and $C^{(3)}$ field then become: 
\begin{align}
\label{Msol3}
	ds_{11}^2=&e^{2A_0}(-dt^2+dy^2)+e^{-A_0}(-\partial_z w)^{-\frac{1}{2}}\left(du_1^2+du_2^2+du_3^2+du_4^2\right)  \\
	&+e^{2A_0}(-\partial_z w)\left(dz+(\partial_z w)^{-1}(\partial_{u_1} w)du_1\right)^2 + e^{-A_0}(-\partial_z w)^{\frac{1}{2}}\left(d\upsilon^2+\upsilon^2 d\Omega_3'^2\right) \,, \nonumber \\
\label{Msol4}
	C^{(3)}=&-e^{3A_0}(-\partial_z w)^{\frac{1}{2}}dt\wedge dy\wedge dz+e^{3A_0}(-\partial_z w)^{-\frac{1}{2}} (\partial_{x_1}w)dt\wedge dy\wedge dx_1  \\
	&+(-\partial_z w)^{-1}(\partial_{u_1} w)du_2 \wedge du_3 \wedge du_4 + (\upsilon^3 \partial_{\upsilon}w)d\Omega_3' \,. \nonumber
\end{align} 
Comparing \eqref{Msol3} with \eqref{M sol1} it is easy to see that the following relations should hold for the two metrics to be the same:
\begin{equation}
	e^{2A_0}=\frac{h_{11}^{1/3}}{(\det h)^{2/3}} \,, \hspace{5pt} h_{11}=-\partial_z w \,, \hspace{5pt} h_{12}=-\partial_{u_1} w \,.
\end{equation}
Finally, comparing \eqref{Msol4} with \eqref{M sol2} we see that $w$ should be equal to $-\frac{1}{2}\partial_z K$, and from  (\ref{solfns}) one finds that the \mname\ function,  $G_0$,  is, up to signs and factors of $2$, precisely the K\"ahler potential of the string-web solution.

\section{Floating M2 and M5 branes}
\label{sec:floatingM2}

Even if the brane structure of the solutions we construct is obscured by the brane interactions and back-reaction, there is a very intuitive way to reveal this structure: one finds the brane probes that feel no force when inserted into these solutions. The orientation of the floating branes in the metric given by the frames in Section \ref{sec:GenRes} is straightforward, and so is the evaluation of the appropriate DBI-like and Wess-Zumino-like terms in the M2-brane action. However, the action of M5 branes is rather complicated \cite{Bandos:1997ui}; 
the easiest way to evaluate it is to use one of the isometries of the M-theory solution to formally write it as a Type IIA solution (as we did in the previous section) and evaluate the action of probe D4 branes.

We now examine these brane probes in more detail.

\subsection{Floating M2 branes in the intersecting M2-M5 Ansatz}
\label{ss:uvzfloat}

It is easy to see the floating M2 branes in the  brane-intersection formulation of the M2-M5 solutions of Section \ref{sec:GenRes}.  Indeed, parametrizing the brane using coordinates $(\eta_0, \eta_1, \eta_2)$, it is trivial to see that a brane with:
\begin{equation}
t  ~=~ \eta_0  \,,  \qquad y  ~=~\eta_1 \,,  \qquad z  ~=~\eta_2 \,, \qquad \vec u, \vec v \ {\rm constant}  \,, 
\label{floatM2}
\end{equation}
feels no force in all of the solutions in Section  \ref{sec:GenRes}.  This is because the $(t,y,z)$-components of $C^{(3)}$ are simply:
\begin{equation}
C_{tyz}^{(3)} ~=~   - e^0 \wedge e^1 \wedge e^2   \,.
 \label{C3tyz}
\end{equation}
Since this is the form of the determinant of the frames along the probe directions, it means that the WZW term will precisely cancel the DBI action for the brane embedding defined by (\ref{floatM2}).  Thus (\ref{floatM2}) defines floating M2 branes.  

Using  (\ref{variables3}) we see that in the AdS$_3$ formulation, the floating M2 branes are given by: 
\begin{equation}
\rho ~=~  k_1 \, \mu^{-1} \,, \qquad \tilde \Phi(\xi,\rho)~=~  k_2  \,,
 \label{AdSfloat}
\end{equation}
where $k_1$ and $k_2$ are constants.  The floating M2 branes thus follow the level-curves of $\tilde \Phi$, with scale, $\mu$, set by the radial coordinate, $\rho$, on the Riemann surface.

\subsection{Floating M2 branes in the \texorpdfstring{AdS$_3$}{AdS3} formulation}
\label{ss:AdSfloat}

It is instructive to look for floating M2 branes directly in the AdS$_3$ solutions.  We are going to start with a general value (though positive) value of $\gamma >0$.  Following (\ref{floatM2}) we use the parametrization
\begin{equation}
t  ~=~ \eta_0  \,,  \qquad y  ~=~\eta_1 \,,  \qquad \mu   ~=~e^{\eta_2} \,, \qquad \xi    ~=~\sigma_1(\eta_2)  \,,\qquad \rho   ~=~\sigma_2(\eta_2) \,, 
\label{floatM2AdS}
\end{equation}
for some functions, $\sigma_1$ and $\sigma_2$.

The pull-back of the metric defined by  (\ref{DHokerAnsatz}),  (\ref{PoincareMet}) and  (\ref{PoinAdS})   onto the M2 brane is:
\begin{equation}
d\hat s_3^2 ~=~    e^{2A} \, \Big[ \, \hat f_1^2 \, \big( d \eta_2^2 ~+~e^{2 \eta_2} \,\big(  -d \eta_0^2 + d \eta_1^2\,\big)    \big) ~+~ \frac{(\sigma_1')^2 + (\sigma_2')^2  }{4\,  \sigma_2^2}\,d \eta_2^2 \, \Big]  \,.
\label{pullbackmet}
\end{equation}
The DBI Lagrangian is given by the square-root of the determinant of this metric:
\begin{equation}
\begin{aligned}
\cL_{DBI} ~=~  &   e^{3A} \,\hat f_1^2\, e^{2 \eta_2} \,  \bigg( \hat f_1^2 + \frac{(\sigma_1')^2 + (\sigma_2')^2  }{4\,  \sigma_2^2}\bigg)^{\frac{1}{2}} \\
 ~=~  &  h\,    \,\hat f_1^2\, e^{2 \eta_2} \,  \bigg[ (\, G    \overline G~-~1)  \, W_+\, W_- \, \bigg(\hat f_1^2 + \frac{(\sigma_1')^2 + (\sigma_2')^2  }{4\,  \sigma_2^2}\bigg)\bigg]^{\frac{1}{2}}  \,.
\end{aligned}
\label{DBILag}
\end{equation}
To be able to cancel this against the WZW term, the term in the square bracket needs to be a perfect square.  For $\gamma=1$, there is a simple way to achieve this.  Suppose  
\begin{equation}
\frac{(\sigma_1')^2 + (\sigma_2')^2  }{ \sigma_2^2}  ~=~ \frac{g_1^2  + g_2^2}{ g_1^2}     \,, 
\label{sigmaforms}
\end{equation}
then one finds that 
\begin{equation}
\bigg[ (\, G    \overline G~-~1)  \, W_+\, W_- \, \bigg(\hat f_1^2 + \frac{(\sigma_1')^2 + (\sigma_2')^2  }{4\,  \sigma_2^2}\bigg)\bigg]      ~=~ \bigg( \frac{W_+ W_-}{4 \,g_1}\bigg)^2   \,,
\label{miracle}
\end{equation}
and the DBI Lagrangian  reduces to:
\begin{equation}
\cL_{DBI}   ~=~  e^{2 \eta_2} \, \frac{\beta \, \sigma_2 \,\big( (g_1^2+g_2^2)^2 - g_2^2 \big)}{4  \,g_1 \, (g_1^2+g_2^2 -1) }  \,,
\label{DBILag-red}
\end{equation}
where we have assumed $\gamma =1$ and used (\ref{simph}) and  (\ref{floatM2AdS}).

The WZW action is the pull-back of $C^{(3)}$ onto the M2 brane:
\begin{equation}
\widehat C^{(3)} ~=~    b_1 \, e^{2 \eta_2} \,  d \eta_0 \wedge  d \eta_1 \wedge d \eta_2 \,, 
\label{pullbackC}
\end{equation}
with $b_1$ given by (\ref{bfunctions}).  However, we must also allow for a possible gauge transformation of the form  $C^{(3)} \to C^{(3)}  + d [\mu^2 \Lambda(\rho,\xi) \, d\eta_0 \wedge d\eta_1]$,  for some function, $\Lambda$, and so we take the WZW action to be:
\begin{equation}
\widetilde C^{(3)} ~=~    e^{2 \eta_2} \,  \big(\, b_1~+~ 2\,\Lambda  ~+~  (\partial_\xi \Lambda) \,  \sigma_1' ~+~  (\partial_\rho \Lambda) \,\sigma_2' \, \big)\, d \eta_0 \wedge  d \eta_1 \wedge d \eta_2 \,.
\label{C3gauged}
\end{equation}
Taking $\gamma =1$ and  $\Lambda = -2 \nu _1  c_1^{-3}   \Phi$ yields
\begin{equation}
\begin{aligned}
\widetilde C^{(3)} ~=~  &   e^{2 \eta_2} \,   \frac{\nu _1}{c_1^3}\, \bigg[\,    \frac{ h \, (G + \overline G) }{(G    \overline G~-~1) }      ~-~  2\, (\partial_\xi \Phi) \,  \sigma_1'  ~-~  2\,(\partial_\rho \Phi) \, \sigma_2' \, \bigg]\, d \eta_0 \wedge  d \eta_1 \wedge d \eta_2 \\
~=~  &  \nu _1\,  e^{2 \eta_2} \,   \frac{\beta\,  \sigma_2 }{4 }\, \bigg[\,    \frac{ g_1 }{(g_1^2+g_2^2 -1)}      ~+~  g_2 \,  \frac{\sigma_1'}{\sigma_2}  ~-~   g_1 \,  \frac{\sigma_2'}{\sigma_2} \, \bigg]\, d \eta_0 \wedge  d \eta_1 \wedge d \eta_2 \,, 
\end{aligned}
\label{C3fixed}
\end{equation}
where we have used  (\ref{Phidefns}) and  (\ref{params}) for $\gamma =1$.  

The WZW term exactly matches the DBI Lagrangian, (\ref{DBILag-red}), if and only if:
\begin{equation}
 \sigma_2 ~=~      k_1  \, e^{-\eta_2} \,,  \qquad \frac{\sigma_1'}{\sigma_2}  ~=~ \frac{g_2}{g_1}\,,
\label{sigeqns}
\end{equation}
which also satisfies (\ref{sigmaforms}).  The constant, $k_1$ is the same as in (\ref{AdSfloat}).

Given that $\sigma_2 = -\sigma_2'$, the second equation can be written:
\begin{equation}
 g_1 \sigma_1'  ~+~ g_2\,  \sigma_2' ~=~  0 \quad \Leftrightarrow \quad   \partial_\xi \tilde \Phi \, \sigma_1'  ~+~   \partial_\rho \tilde \Phi \, \sigma_2' ~=~  0  \,,
\end{equation}
which means that the floating branes follow the level curves of $\tilde \Phi$. This agrees with the shape  of the floating M2 brane determined directly in Section \ref{ss:uvzfloat} and changing from the M2-M5 coordinates to the coordinates proper to the AdS$_3$ solution  (\ref{AdSfloat}).

This calculation indicates that floating M2 branes do not exist in the AdS$_3$ solutions of \cite{Bachas:2013vza} unless  $\gamma =1$. Only for this value the expression under the square root in the DBI Lagrangian  becomes a perfect square, allowing the DBI and the WZ Lagrangians to cancel. This confirms the result of the previous section: the AdS$_3$ solutions of \cite{Bachas:2013vza} only correspond to near-horizon limits of M2-M5 brane intersections in flat space when $\gamma=1$.

\subsection{Floating M5 branes in the intersecting M2-M5 Ansatz}

As we mentioned above, evaluating the action of probe M5 branes in a complicated background is quite involved. The easiest strategy is to  reduce the solution to Type IIA and evaluate the action of probe D4 branes. Fortunately, we have already obtained the formulas for the IIA reduction of our M2-M5 solution when relating it to F1-D1 string webs \eqref{D2-D4 final}.

We consider probe D4 branes whose volume is parametrized by $(\eta_0, \eta_1, \eta_2, \eta_3, \eta_4)$ embedded in spacetime as 
\begin{equation}
\label{floating D4}
	\eta_0=t\,, \hspace{5pt} \eta_1=-u_1\,, \hspace{5pt} \eta_2=u_2 \,, \hspace{5pt} \eta_3=u_3\,, \hspace{5pt} \eta_4=y \,.
\end{equation}
The induced metric on it is
\begin{equation}
\label{induced D4}
	d\tilde{s}_5^2=\frac{1}{\sqrt{\det h}}(-dt^2+dy^2)+\frac{h_{22}}{\sqrt{\det h}}du_1^2+\frac{\sqrt{\det h}}{h_{11}}(du_2^2+du_3^2)
\end{equation}
and the induced NS-NS and RR fields are
\begin{align}
\label{ind D4 flux}
	\tilde{B}_2&=\frac{h_{12}}{h_{11}}du_2\wedge du_3 \,, \nonumber  \\
	\tilde{C}_3&=-\frac{h_{12}}{\det h}dt\wedge du_1 \wedge dy\,, \\
	\tilde{C}_5&=-\frac{1}{h_{11}}dt\wedge du_1 \wedge du_2 \wedge du_3 \wedge dy\,. 	                                                                                                                                       \nonumber
\end{align}
It is straightforward now to see that the DBI and WZ actions are:
\begin{align}
	S_{DBI}&=-T_4\int\,d^5\sigma e^{-\phi}\sqrt{-\det \left( \tilde{G}_{\alpha \beta}+F_{\alpha \beta}+\tilde{B}_{\alpha \beta}\right)} = -T_4 \int \, d^5\sigma \frac{h_{22}}{\det h} \,, \\
	S_{WZ}&=-T_4 \int\, e^{\tilde{B}_2+\tilde{F}_2}\wedge \oplus_n \tilde{C}_n=T_4 \int d^5\sigma \frac{h_{22}}{\det h}\,,
\end{align}
and hence the D4-brane \eqref{floating D4} feels no force in this solution. This in turn indicates that probe M5 branes extended along $y$ and the four-torus parameterized by $\vec u$ feel no force in the M2-M5 solution (\ref{11metric}-\ref{C3gen}).


\section{Interesting examples of M2-M5 near-horizon solutions}
\label{sec:examples}

The solutions of the M2-M5 system considered in Section \ref{sec:GenRes} are determined by a non-linear Monge-Amp\`ere-like \mname\ equation (\ref{master}), and obtaining generic solutions to this equation is prohibitively complicated. Even simpler solutions such as a single infinite M2 brane ending on (and pulling on) an M5 brane cannot be found. The only known solution to this \mname\ equation is the one corresponding to an infinite tilted M5 brane with M2 charge. This solution can be obtained by dualizing a tilted D-brane system, and for the proper tilt and the proper M2 brane density it can be shown to fit precisely in the Ansatz (\ref{11metric-symm}). We  present this solution in Appendix~\ref{app:tiltedD2}.

In contrast, the near-horizon geometries corresponding to M2-M5 solutions with an $SO(4) \times SO(4)$ isometry have been shown in Section \ref{sec:AdSlmits} to belong to the  $\gamma=1$ family of 
AdS$_3 \times$S$^3 \times$S$^3 \times \Sigma$ solutions constructed in  \cite{Bachas:2013vza}, where $\Sigma$  is a Riemann surface. These solutions can be constructed systematically by solving a set of linear equations in two dimensions, and there are quite a few classes of solutions that one can try to relate to M2-M5 brane systems. 

\subsection{No-flip solutions}
\label{ss:form}

In Section 8.3 of \cite{Bachas:2013vza}, the authors consider ``no-flip''  solutions with:
\begin{equation}
h  ~=~ -i w + i \bar w \,,  \qquad G ~=~ \pm \Bigg[  \, i  ~+~  \sum_{a=1}^{n+1}  \frac{\zeta_a \,{\rm Im}(w)}{(\bar w -\xi_a) \vert w - \xi_a\vert} \,  \Bigg] \,, 
\label{Gintsols}
\end{equation}
where $\xi_a$ and $\zeta_a$ are real parameters. As noted in (\ref{interchanges}), the exchange of the plus and minus sign in the expression above exchanges the role of $u$ and $v$ and the role of $z$ and $w$. 

From  (\ref{Gintsols}), (\ref{wparts}) and (\ref{simph}), we see that
\begin{equation}
\beta  ~=~ 2 \,, 
\label{betaval}
\end{equation}
and 
\begin{equation}
g_1 ~=~ \pm \, \sum_{a=1}^{n+1}  \, \frac{\zeta_a \, \rho \, (\xi - \xi_a)}{  \big( (\xi - \xi_a)^2 + \rho^2\big)^\frac{3}{2}}  \,, \qquad g_2~=~ \pm \bigg[  \, 1 ~+~   \sum_{a=1}^{n+1}  \, \frac{\zeta_a \,  \rho^2}{  \big( (\xi - \xi_a)^2 + \rho^2\big)^\frac{3}{2}}  \,  \bigg]  \,.
\label{ReGImG}
\end{equation}

Such functions are familiar in the study of axi-symmetric Gibbons-Hawking metrics\footnote{This may not be a complete coincidence: Gibbons-Hawking metrics provide solutions to the Monge-Amp\`ere equation in two variables with two $U(1)$ isometries.}.  Indeed, one can easily compute the potentials from (\ref{Phidefns}) (with $\beta =2$):
\begin{equation}
\tilde \Phi ~=~ \pm \,2\, \bigg[ \,  - \log \rho~+~ \sum_{a=1}^{n+1}  \, \frac{\zeta_a }{   \sqrt{ (\xi - \xi_a)^2 + \rho^2} } \,\bigg] \,, \qquad  \Phi ~=~ \mp  2\,  \bigg[  \,  \xi ~+~ \sum_{a=1}^{n+1}  \, \frac{\zeta_a \, (\xi - \xi_a)}{   \sqrt{ (\xi - \xi_a)^2 + \rho^2} }  \,  \bigg]  \,.
\label{Phis-example}
\end{equation}
This gives a nice physical picture of these solutions: the non-logarithmic part of $\tilde \Phi$ is simply the three-dimensional potential of charges, $\pm 2\, \zeta_a$, arrayed along the $\xi$-axis at the points $(\rho,\xi) = (0,\xi_a)$. The $\log \rho$ component of $\tilde \Phi$ is also a harmonic function, and its role is to give the correct asymptotic behavior.  
\begin{figure}[ht]
  \centering
  \includegraphics[width=.6\linewidth]{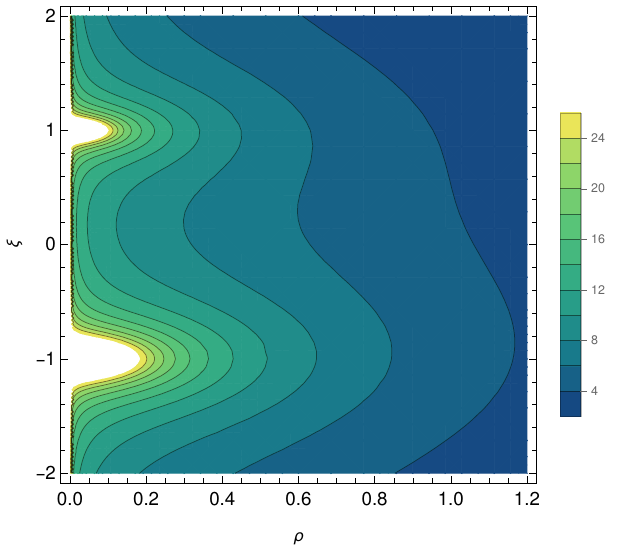}
  \caption{Contour plot of $\tilde\Phi$ for two points, with $(\zeta_1 = 1, \xi_1 = 1)$ and $(\zeta_2 = 2, \xi_2 = -1)$. By \eqref{AdSfloat}, the lines of constant potential drawn in black are the floating M2 branes.}
  \label{M2-profile}
\end{figure}

The floating M2 branes are given by  (\ref{AdSfloat}).  
When $|k_2|$ is large, then either  $u$ or $v$ must be small. The floating branes are thus given by the level-curves of $\tilde \Phi$  and, as can be seen in Fig.~\ref{M2-profile}, when $\rho$ is small, they  roughly circle  half-way around each of of the  singularities of $\tilde \Phi$ located at  
$(\rho,\xi) = (0,\xi_a)$, and run roughly parallel to the boundary  between the singularities. These infinite M2 branes are parallel to the M2 branes whose back-reaction gives rise to the solution, but in the coordinates adapted to the $AdS_3$ near-horizon geometry they have the shape given in Figure \ref{M2-profile}.

\section{Adding momentum}
\label{sec:Mom}
To transform the M2-M5 supermaze we discussed in the previous sections into a microstate of a three-charge black hole with a large horizon, one needs to add to it momentum along the common M2-M5 direction. As discussed in detail in \cite{Bena:2022wpl,Bena:2022fzf}, a momentum wave that preserves the locally-16-supercharge structure of the supermaze can only be added when accompanied by other brane dipole moments, which can modify the structure of our Ansatz. Our purpose here is to find the minimal modification of our Ansatz needed to add momentum, and to construct the resulting solution, even though this solution does not display local supersymmetry enhancement to 16 supercharges.

\subsection{Adding momentum to spherically symmetric brane intersections}
\label{ss:mom-iint}

As we will see, the simplest \nBPS{8} solutions that carry momentum charge have a singular momentum-charge source that gives rise to a momentum harmonic function that encodes the momentum density of the solution. 

We start from the metric Ansatz:
\begin{equation}
\begin{aligned}
ds_{11} =- e^{2  A_0}\, dt^2 ~+~ e^{2  A_1}\,  \big( dy  - P\, dt \big)^2   
& ~+~ e^{2A_2} \, du^2 ~+~ e^{2A_3} \, dv^2  ~+~ u^2 \,  e^{2A_4} \, d\Omega_3^2    ~+~ v^2 \,  e^{2A_5} \, d{\Omega'}_3^2  \\ &~+~ e^{2  A_6}\, \big( dz ~+~B_1  \, d u  \big)^2     \,,
\end{aligned}
 \label{11metric-mom}
\end{equation}
%
%
%
which reduces to the no-momentum Ansatz in equation \eqref{11metric} by taking $A_1 \equiv A_0$ and $P \equiv 0$.  We  take the frames to be:
\begin{equation}
\begin{aligned}
e^0 ~=~  & e^{A_0}\, d  t \,, \qquad e^1~=~  e^{A_1}\,  \big( dy  - P\, dt \big) \,,  \qquad e^2 ~=~    e^{A_6}\, \big( dz ~+~B_1  \, d u \big) \,,\\
   e^3 ~=~ &  e^{A_2} \, du\,,  \qquad e^4 ~=~   e^{A_3} \, dv  \,, \qquad e^{i+4}~=~ u\,  e^{A_4} \,  \sigma_i  \,, \qquad e^{i+7}  ~=~  v\, e^{A_5} \,  \tilde \sigma_i \,,   \qquad {i = 1,2,3}   \,.
\end{aligned}
 \label{frames-mom}
\end{equation}
Based on the symmetries, we also use the Ansatz (\ref{F4gen}) for the field strength.

 The supersymmetries of this system will be defined in terms of the frame components along the momentum direction, $y$,  as well as the M2 and M5 directions: 
\begin{equation}
 \Gamma^{01} \, \varepsilon  ~=~ - \varepsilon \,,   \qquad   \Gamma^{012} \, \varepsilon  ~=~ - \varepsilon \,,   \qquad  \Gamma^{013456} \, \varepsilon  ~=~\varepsilon 
 \label{projs-new}
\end{equation}
There are thus four supersymmetries and the brane system is \nBPS{8}.

As before, these projectors are compatible with 
\begin{equation}
 \Gamma^{01789\,1\!0} \, \varepsilon  ~=~-\varepsilon \,,
 \label{projs2a}
\end{equation}
and hence one can add another set of M5 branes along the directions $01789\,10$ without breaking supersymmetry any further.   

As before, the goal is  to solve
\begin{equation}
\delta \psi_\mu ~\equiv~ \nabla_\mu \, \epsilon ~+~ \coeff{1}{288}\,
\Big({\Gamma_\mu}^{\nu \rho \lambda \sigma} ~-~ 8\, \delta_\mu^\nu  \, 
\Gamma^{\rho \lambda \sigma} \Big)\, F_{\nu \rho \lambda \sigma}\,\epsilon ~=~ 0 \,,
\label{11dgravvar-repeat}
\end{equation}
subject to the foregoing projection conditions.

Solving these BPS equations proceeds as in Appendix \ref{ss:BPSsystem}, except that one rapidly discovers that one must take $P = - e^{A_0 - A_1}  + const$.  The constant can be absorbed through a shift of $y$, and we will take: 
\begin{equation}
P   ~\equiv~1 ~-~  e^{A_0 - A_1}\,.
 \label{Psol}
\end{equation}
With this choice, the solution with no momentum constructed in Appendix \ref{ss:BPSsystem} is simply given by taking $A_1 = A_0$.  

It is convenient to define 
\begin{equation}
\hat A_0    ~\equiv~  \frac{1}{2}\, (A_0 +  A_1)  \,,  \qquad \hat A_1    ~\equiv~ \frac{1}{2}\, (A_0 -  A_1)  \,.
 \label{hatAdefns}
\end{equation}
and one finds that all the remaining BPS equations are {\it exactly} as in Section \ref{ss:BPSsystem} with $\hat A_0$ replacing by $A_0$ and $A_6$ replacing $A_1$.  In particular,  the BPS equations place   no constraint whatsoever on $\hat A_1$.

One thus finds that the BPS equations are identically satisfied if the metric takes the form
\begin{equation}
\begin{aligned}
ds_{11}  = e^{2  \hat A_0}\, \bigg[&  -  e^{2  \hat A_1} \, dt^2  +  e^{- 2  \hat A_1} \,    \Big( dy  +\big (e^{2\,\hat A_1} -1\big) \, dt \Big) ^2  +  (-\partial_z w ) \, \Big( dz ~+~(\partial_z w )^{-1}\,  \big (\partial_u w \big)  \, d u \Big)^2   \\
& ~+~ e^{-3  \hat A_0} \, (-\partial_z w )^{-\frac{1}{2}}\, \big( du^2  ~+~ u^2 \, d\Omega_3^2 \big)   ~+~ e^{-3  \hat A_0} \, (-\partial_z w )^{\frac{1}{2}}\, \big( dv^2  ~+~ v^2 \, d{\Omega'}_3^2 \big)  
  \bigg]\,,
\end{aligned}
 \label{11metric-symm-new}
\end{equation}
with the frames:
\begin{equation}
\begin{aligned}
e^0 ~=~  &e^{\hat A_0 + \hat A_1} \, dt \,, \qquad e^1~=~ e^{\hat A_0- \hat A_1}\,  \Big( dy  +\big (e^{2\,\hat A_1} -1\big) \, dt \Big) \,,  \\  e^2 ~=~ &   e^{\hat A_0} \, (-\partial_z w )^{\frac{1}{2}}\,  \Big( dz ~+~(\partial_z w )^{-1}\,  \big (\partial_u w \big)  \, d u \Big) \,,\\ 
e^3 ~=~  & e^{-\frac{1}{2} \hat A_0} \, (-\partial_z w )^{-\frac{1}{4}}\,   du\,,  \qquad  e^4 ~=~  e^{-\frac{1}{2} \hat A_0} \, (-\partial_z w )^{\frac{1}{4}}\,   \, dv  \,, \\   e^{i+4}~=~ & \frac{1}{2} \, u\,   e^{-\frac{1}{2} \hat A_0} \, (-\partial_z w )^{-\frac{1}{4}} \,  \sigma_i  \,, \qquad e^{i+7}  ~=~   \frac{1}{2} \, v\,   e^{-\frac{1}{2} \hat A_0} \, (-\partial_z w )^{\frac{1}{4}} \,   \tilde \sigma_i \,,   \qquad {i = 1,2,3}   \,.
\end{aligned}
 \label{RotInvframes-new}
\end{equation}
One finds that $C^{(3)}$ is still given by given by (\ref{C3symm}): 
\begin{equation}
C^{(3)} ~=~   - e^0 \wedge e^1 \wedge e^2 ~+~  (\partial_z w )^{-1} \, \big (u^3 \partial_u w \big)  \, {\rm Vol}({S^3}) ~+~  \, \big(v^3 \partial_v w \big) \,  {\rm Vol}({{S'}^3})    \,.
 \label{C3symm-repeat}
\end{equation}
Note that $\hat A_1$ cancels out entirely in $e^0 \wedge e^1$.  Thus the solution to the BPS equations is exactly as it was in the absence of momentum, except for the $\hat A_1$-dependent terms  in the metric.  

As before the BPS solution is obtained by solving   (\ref{MALap}) and then $w$ and $\hat  A_0$ are obtained from (\ref{G0defn}) and the appropriate re-labeling of (\ref{F12defns}):
\begin{equation}
F_1   ~\equiv~ (-\partial_z  w)^{\frac{1}{2}} \,e^{-3   \hat A_0} \,, \qquad  F_2  ~\equiv~ (-\partial_z  w)^{-\frac{1}{2}} \,e^{-3   \hat A_0}~+~  (-\partial_z  w)^{-1} \,  (\partial_u w)^2\,,
 \label{F12defns-new}
\end{equation}

The function $\hat A_1$ is not determined by the BPS equations, however it is determined by the equations of motion.  To that end, it is useful to define the operator, ${\cal L}$, to be the Laplacian of the metric (\ref{11metric-symm-new}) with $\hat A_1 =0$.  If $H(u,v,z)$, is only a function of $(u,v,z)$, then one finds that the equations for $w$ and $\hat A_0$ enable one to simplify the Laplacian  to:
\begin{equation}
\begin{aligned}
{\cal L} ( H)    ~=~    e^{2  \hat A_0}\,(-\partial_z w )^{-\frac{1}{2}} \,   \bigg[&  (- \partial_z w) \, \frac{1}{u^3} \partial_u \big( u^3  \partial_u H \big) ~+~  \frac{1}{v^3} \partial_v \big( v^3  \partial_v H \big) ~+~2\, ( \partial_u w) \partial_u \partial_z H     \\
  & ~+~  \Big((-\partial_z  w)^{-\frac{1}{2}} \,e^{-3   \hat A_0}~+~  (-\partial_z  w)^{-1} \,  (\partial_u w)^2\big)\Big)\, \partial_z^2 H \bigg] \,.
\end{aligned}
 \label{Lap}
\end{equation}
Note that, by definition (and as is manifest from the explicit expression), this is a linear operator on the background geometry defined by $\hat A_0$ and $w$. 

One can show that all the equations of motion are satisfied if $\hat A_1$ solves:
\begin{equation}
{\cal L} \big( e^{-2  \hat A_1}\big)     ~=~ 0\,.
 \label{harmonicP}
\end{equation}
Thus there is a simple harmonic Ansatz for adding pure momentum sources to the the M2-M5 brane intersections.

\subsection{Adding momentum charge to the near-brane limit}
\label{ss:mom-near-brane}

From the results above, it is elementary to translate the effect of adding a source corresponding to momentum to the AdS near-brane limit described in Section \ref{sec:AdSlmits}.  The metric given by (\ref{DHokerAnsatz}),  (\ref{PoincareMet}) and (\ref{PoinAdS}) generalizes to:
\begin{equation}
\begin{aligned}
ds_{11}^2 ~=~   e^{2A} \, \bigg[ & \, \hat f_1^2 \,    \bigg( \frac{d \mu^2}{\mu^2} ~+~ \mu^2 \, \Big( -  e^{2  \hat A_1} \, dt^2  +  e^{- 2  \hat A_1} \,    \big( dy  +\big (e^{2\,\hat A_1} -1\big) \, dt \big) ^2  \Big) \bigg)  \\ & ~+~ \hat f_2^2 \, ds_{S^3}^2 ~+~ \hat f_3^2 \, ds_{{S'}^3}^2 ~+~ \frac{ d\xi^2  + d\rho^2}{4  \rho^2}   \,   \bigg] \,,
\end{aligned}
 \label{near-brane-mom}
\end{equation}
and the flux, $C^{(3)}$, remains the same.  

The function, $\hat A_1$, is determined by the harmonic condition,  (\ref{harmonicP}), and in the coordinate system of the near-brane $AdS$ limit of Section \ref{sec:AdSlmits}, this Laplacian becomes: 
\begin{equation}
\begin{aligned}
{\cal L} ( H)    ~=~  4\, e^{-A} \, \bigg[\,(G \bar G -1)\, \frac{1}{\mu}\, \partial_\mu \big( \mu^3 \partial_\mu H \big)  ~+~\frac{1}{\rho}\, \partial_\rho \big( \rho^3 \partial_\rho H \big) ~+~ \rho^2 \partial_\xi^2 H \, \bigg] \,,
\end{aligned}
 \label{Lap-AdS}
\end{equation}
on some function, $H(\mu, \xi, \rho)$. 

If one seeks a scaling solution to ${\cal L} ( H) =0$ with  $H(\mu, \xi, \rho) = \mu^p K(\xi,\rho)$, then $K$ must satisfy
\begin{equation}
\begin{aligned}
\frac{1}{\rho}\, \partial_\rho \big( \rho^3 \partial_\rho K \big) ~+~ \rho^2 \partial_\xi^2 K~+~ p(p+2)\,  (G \bar G -1)\, K  ~=~  0\,.
\end{aligned}
 \label{redLap-AdS}
\end{equation}
 One can easily verify that, for $p=-1$, this has solutions
\begin{equation}
\begin{aligned}
K  ~=~  \frac{c_1}{u^2} ~+~   \frac{c_2}{v^2} ~=~  \frac{2}{\beta \rho} \big(c _1 \, e^{\frac{1}{\beta} \tilde \Phi} + c _2 \, e^{-\frac{1}{\beta} \tilde \Phi}\big) \,.
\end{aligned}
 \label{samplesols}
\end{equation}
where $c _1$ and $c _2$ are constants and we have used (\ref{variables3}).  This is also manifestly a solution to ${\cal L} ( H) =0$ using  (\ref{Lap}).  These correspond to smeared distributions of singular sources of the momentum harmonic function, 

To get an isolated momentum source one would like a solution that falls off at spatial infinity as $(u^2 +v^2)^{-3}$. In the $\mu, \rho$ coordinates this is
\begin{equation}
(u^2 +v^2)^{-3}  ~\sim~ \mu^{-3} \, \rho^{-3}  \,.
 \label{asympfalloff}
\end{equation}

It is useful to note that  (\ref{redLap-AdS}) may be written as 
\begin{equation}
\frac{1}{\rho^3 }\, \partial_\rho \big( \rho^3 \partial_\rho K \big) ~+~ \partial_\xi^2 K~+~ \frac{p(p+2)}{\rho^2}\, (G \bar G -1)\, K  ~=~  0\,, 
 \label{redLap-eqn}
\end{equation}
and that 
\begin{equation}
{\cal L}_4 (K) ~\equiv~  \frac{1}{\rho^3 }\, \partial_\rho \big( \rho^3 \partial_\rho K \big) ~+~ \partial_\xi^2 K \,, 
 \label{L4defin}
\end{equation}
is the Laplacian of flat $\IR^5$ with a metric: 
\begin{equation}
\begin{aligned}
ds_5^2 ~\equiv~ d\rho^2 ~+~ \rho^2 d\Omega_3^2 ~+~  d \xi^2\,.
\end{aligned}
 \label{fivemet}
\end{equation}
This means that 
\begin{equation}
{\cal L}_4 \bigg(  \frac{1}{(\rho^2 + \xi^2)^{\frac{3}{2}}}\bigg) ~=~ 0 \,.
 \label{homsol}
\end{equation}
Note that this falls off as $\rho^{-3}$ at large $\rho$, which is consistent with (\ref{asympfalloff}).

Now observe that at large $\rho$ and $\xi$, one generically has $(G \bar G -1) \to 0$.  Indeed, the example in Section \ref{ss:form} one has:
\begin{equation}
\frac{p(p+2)}{\rho^2}\, (G \bar G -1)  ~\sim~ \frac{c_0}{(\rho^2 + \xi^2)^{\frac{3}{2}}} \,, 
 \label{asympGbarG}
\end{equation}
for some constant $c_0$.  One can then solve (\ref{redLap-AdS}) perturbatively.  Indeed, if one starts from the homogeneous solution (\ref{homsol}), then, using  (\ref{asympGbarG}), one obtains, at first order:
\begin{equation}
K  ~=~ \frac{Q}{(\rho^2 + \xi^2)^{\frac{3}{2}}}\,  \bigg( 1 ~-~\frac{c_0}{4}\, \frac{1}{\sqrt{\rho^2 + \xi^2} } ~+~ \dots \bigg) \,, 
 \label{Kasymp}
\end{equation}
for some constant, $Q$, that determines the momentum charge.  Given the simplicity of the Laplacian and the forms of the solutions in Section \ref{ss:form}, this perturbation expansion can be continued to arbitrarily high order.

\subsection{An interesting family of solutions }
\label{ss:CoolExample}

A simple solution to (\ref{Lap-AdS}) is:
\begin{equation}
e^{-2  \hat A_1}  ~=~ V_0 ~+~ V_1 \, \mu^{-2}  \,,
\label{Ahatsol1}
\end{equation}
where
\begin{equation}
V_0 ~=~  1~+~ \sum_{a=1}^{m}  \, \frac{k_a }{  \big( (\xi -\tilde \xi_a)^2 + \rho^2\big)^\frac{3}{2}}  \,, \qquad V_1 ~=~   q_0 ~+~   \sum_{a=1}^{m'}  \, \frac{q_a }{  \big( (\xi - \hat \xi_a)^2 + \rho^2\big)^\frac{3}{2}}  \,, 
\label{V0V1defns}
\end{equation}
for some charges $k_a$ and $q_a$. Note that we have taken the constant term in  $V_0$ to be $1$ because we require $ \hat A_1 \to 0$ at infinity.  Furthermore, the locations of the poles of these harmonic functions,  $\tilde \xi_a$ and $\hat \xi_a$ do not have to coincide with $\xi_a$, the locations of the poles of $\Phi$ and $\tilde \Phi$ in Section \ref{ss:form}, but it might be interesting if they did.

Amusingly enough, at fixed $\rho$ and $\xi$, (\ref{Ahatsol1}) gives 
\begin{equation}
e^{-2  \hat A_1}  ~=~ 1 + \alpha ~+~Q \, \mu^{-2}  \,,
\label{Ahatsol2}
\end{equation}
for some parameters, $\alpha, Q >0$.

If one takes $k_a =0, a \ge 1$, then $\alpha =0$ and the metric in the AdS$_3$  directions becomes
\begin{equation}
\begin{aligned}
 &  \frac{d \mu^2}{\mu^2} ~+~ \mu^2 \, \Big( -  e^{2  \hat A_1} \, dt^2  +  e^{- 2  \hat A_1} \,    \big( dy  +\big (e^{2\,\hat A_1} -1\big) \, dt \big) ^2  \Big) \\
  &~=~  \frac{d \mu^2}{\mu^2} ~-~   \frac{\mu^4}{Q + \mu^2} \, dt^2  ~+~   (Q + \mu^2)\,    \bigg( dy -  \frac{Q}{Q + \mu^2}  \, dt \bigg) ^2  \\
    &~=~  \frac{d \mu^2}{\mu^2} ~+~ \mu^2 \big(  -  dt^2  ~+~  dy^2 \big)  ~+~ Q \,( dy -   dt )^2      \,,
\end{aligned}
 \label{BTZlimit}
\end{equation}
which is simply the extremal BTZ metric. Thus, our solution describes a continuous family of extremal BTZ 
$ \times$S$^3 \times$S$^3$ solutions warped over a Riemann surface, where the (angular) momentum of the extremal BTZ solutions, $Q$, is a function of the Riemann-surface coordinates, $\rho$ and $\xi$. The families of solutions we have constructed give an infinite violation of black hole uniqueness with our particular AdS$_3 \times$S$^3 \times$S$^3 \times \Sigma$ asymptotics.

\section{Conclusions and future directions} 
\label{sec:Concl}

We have constructed, from first principles, the eight-supercharge supergravity solutions corresponding to a system of parallel M5 branes with M2 stretched between them, and have related our solutions to those previously obtained in \cite{Lunin:2007mj}. We have found that these solutions are entirely determined by a single ``\mname\ function'' satisfying a Monge-Amp\`ere-like ``\mname\  equation.'' We  have used floating probe M2 and M5 branes to explore the structure of these solutions and have  related a class of our solutions to F1-D1 (p,q) string-web solutions \cite{Lunin:2008tf}.

Solving the \mname\ equation is non-trivial in general, but we have identified two ways of finding special classes of solutions. The first, is to consider an infinite M2-M5 bound state at a certain angle and with a certain ratio of M2 and M5 densities. This solution, obtained in Appendix \ref{app:tiltedD2} by dualities, can be shown to have a \mname\ function that satisfies exactly the \mname\ equation.

The second method to solve this equation is to take a near-horizon limit of our solutions, by imposing a certain scaling symmetry on the functions in the metric and 4-form field-strength Ansatz. This scaling symmetry allowed us, in Section \ref{sec:AdSlmits}, to relate our solutions to a family of the 
AdS$_3 \times$S$^3 \times$S$^3$ solutions warped over a Riemann surface constructed in \cite{Bachas:2013vza}.  As with other microstate geometries, these  solutions can be constructed via a linear procedure. We have discussed the physics of one such family of solutions in this paper, and leave the construction and exploration of more general solutions to future work.

We have also found a family of supergravity solutions that describe M2-M5 intersections carrying momentum. The momentum of these solutions has singular sources, but we have succeeded in extending the solutions of \cite{Bachas:2013vza} to BTZ$^{\rm extremal} \times$S$^3 \times$S$^3$ solutions warped over a Riemann surface, where the momentum of the BTZ black hole is a function of the Riemann surface coordinates.
From a higher-dimensional perspective, these solutions violate black-hole uniqueness copiously.

The primary focus of our work here has been  the construction of  the ``static,'' or momentum-free supermazes.  The important next step is to add momentum charge in such a manner that  one obtains themelia \cite{Bena:2022fzf}:  fundamental brane systems that, while being \nBPS{8} globally, actually  have 16 supersymmetries locally, and thus represent  states in the black-hole microstructure.  As we remarked earlier, this will require the addition of  fluxes and localized momentum excitations that go well beyond the simple Ans\"atze we have used here.  On the other hand, our results in Section \ref{sec:Mom} give us considerable optimism that this can be achieved, and perhaps through a linear system of equations that supplements the \mname\ equation.

To understand and appreciate this remark, it useful to recall some of the history of microstate geometries and superstrata.  In the earliest work on microstate geometries, it was clear that the most general such geometry in five dimensions would be based on a general four-dimensional ambi-polar hyper-K\"ahler geometry \cite{Bena:2004de,Bena:2005va,Bena:2007kg}.  Similarly, in six-dimensions, the most general superstrata are based on a highly-non-trivial five-dimensional spatial fibration  over a four-dimensional ``almost hyper-K\"ahler'' base \cite{Gutowski:2003rg,Bena:2011dd}.  These geometries are generically determined by non-linear systems of equations.  However, once these geometries are determined, one can add momentum charge to these backgrounds in a variety of ways, and  the system of equations that determines the momentum excitations, as well as the entire phases space of such excitations,  is actually {\it linear} \cite{Bena:2004de,Bena:2007kg,Bena:2011dd,Giusto:2013rxa,Ceplak:2022wri}. Moreover, the momentum can be added in such a manner as to make themelia, like the superstratum, and this also enables the detailed construction of the corresponding holographic dictionary  \cite{Kanitscheider:2006zf,Kanitscheider:2007wq,Taylor:2007hs,Giusto:2015dfa,Bombini:2017sge,Giusto:2019qig, Tormo:2019yus, Rawash:2021pik}.  

If this pattern repeats with supermazes, then the  ``static,'' or momentum-free supermazes will indeed be governed by generically non-linear \mname\ equations, as we have described here, but it is quite possible that the addition of momentum excitations on top of this geometry could, once again, be governed by linear systems of equations. If this happens, we should be able to add momentum while preserving the 16-local-supercharge structure of the supermaze and thus construct  huge new families of themelia.  Our results in Section \ref{sec:Mom}  are a first step towards achieving of this ideal.

While the linear systems on the ``static,'' or momentum-free supermazes could still be highly non-trivial and difficult to solve explicitly, these structures would still establish the existence of momentum-carrying supermazes in supergravity, and would provide a route to characterizing the phase space of such excitations, and especially the themelia that locally have sixteen supercharges.  Ultimately, one would  like to quantize this phase space to arrive at a semi-classical description of the fractionated branes that lie at the heart of black-hole microstructure.

This may seem like something of ``an ask,'' but the nearly 20 year history of successes in microstate geometries suggests that such a miraculous outcome is really very plausible!

\vspace{1em}\noindent {\bf Acknowledgements:} 
We would like to thank Emil Martinec for  drawing our attention to the work in \cite{Bachas:2013vza} and  its probable relevance to the supergravity description of the supermaze.   We would like to thank Nejc \v Ceplak, Soumangsu Chakraborty and Shaun Hampton for interesting discussions.
The work of IB, AH and NPW was supported in part by the ERC Grant 787320 - QBH Structure. The work of IB was also supported in part by the ERC Grant 772408 - Stringlandscape. The work of AH was also supported in part by a grant from the Swiss National Science Foundation, as well as via the NCCR SwissMAP. The work of DT was supported in part by the Onassis Foundation - Scholarship ID: F ZN 078-2/2023-2024 and by an educational grant from the A. G. Leventis Foundation. The work of NPW was also supported in part by the DOE grant DE-SC0011687.
\newpage

\appendix
\vspace{1em}\noindent {\bf  \LARGE Appendices} 

\section{Spherically symmetric \texorpdfstring{\nBPS{4}}{1/4-BPS} M5-M2 intersections}
\label{app:SphericalSols}

Here we show that our spherically symmetric configurations are, in fact, the most \nBPS{4} general solutions with such symmetries and with supersymmetries defined by (\ref{projs1}).  That is, we will impose Poincar\'e symmetry on the common $(t, y)$ directions of the branes and require an $SO(4) \times SO(4)$ symmetry that sweeps out two three spheres.  We will write down the most general configurations that satisfies these symmetry requirements and, following the methodology developed in \cite{Gowdigere:2002uk,Pilch:2003jg,Gowdigere:2003jf,Nemeschansky:2004yh,Pilch:2004yg}, we will show that  the solutions defined in Section  \ref{ss:sphsymm} are the only possibilities.

\subsection{The Ansatz}
\label{ss:Ansatz}

The most general metric satisfying the symmetry requirements must have the form:
\begin{equation}
\begin{aligned}
ds_{11}^2 ~=~ e^{2  \alpha_0}\, \big(- dt^2 ~+~ dy^2\big)    ~+~  e^{2 \alpha_1} \, d\Omega_3^2    ~+~ e^{2 \alpha_2} \, d{\Omega'}_3^2  ~+~ g_{ij} \,dz^i dz^j     \,,
\end{aligned}
 \label{11metric-Gensymm}
\end{equation}
where $\alpha_0, \alpha_1$ and $\alpha_2$ are arbitrary functions of three remaining coordinates, $z^i$, and $g_{ij}$ is a general metric in these three dimensions.  There is, of course, a remaining diffeomorphism invariance, $z^i \to \tilde z^i (z^j)$, and this can, in principle, be fixed by taking the metric, $g_{ij}$, to be diagonal \cite{DeTurck:1984,Tod:1992,KOWALSKI2013251}.   It is therefore tempting to write $(z^1,z_2,z_3) = (z,u,v)$ and take:
\begin{equation}
\begin{aligned}
ds_{3}^2 ~=~ g_{ij} \,dz^i dz^j  ~=~  e^{2  \alpha_3}\, dz^2    ~+~   e^{2  \alpha_4}\, du^2   ~+~  e^{2  \alpha_5}\, dv^2     \,.
\end{aligned}
 \label{3metric}
\end{equation}
One is then tempted to use a set of frames:
\begin{equation}
\begin{aligned}
e^0 ~=~  & e^{\alpha_0}\, dt \,, \qquad e^1~=~  e^{\alpha_0}\, dy \,,  \qquad e^2 ~=~    e^{\alpha_1}\, dz \, \qquad e^3 ~=~    e^{\alpha_2}\, du \,,\qquad e^4~=~    e^{\alpha_3}\, dv  \,,\\
e^{i+4}~=~& e^{\alpha_4}\,  \sigma_i  \,, \qquad e^{i+7}  ~=~ e^{\alpha_5}\,   \tilde \sigma_i \,,   \qquad {i = 1,2,3}   \,,
\end{aligned}
 \label{genframes}
\end{equation}
however, this misses a very important physical point.  The choice of frames also fixes the meaning of the supersymmetry projectors of the form (\ref{projs1}) and (\ref{projs2}), which in the current frame labelling become:
\begin{equation}
 \Gamma^{012} \, \varepsilon  ~=~- \varepsilon \,,   \qquad  \Gamma^{013567} \, \varepsilon  ~=~\varepsilon \,,   \qquad  \Gamma^{01489\,10} \, \varepsilon  ~=~- \varepsilon  \,.
 \label{projs3}
\end{equation}
The M5, M5' and M2 branes are thereby required to follow the coordinate axes, and this is not the most general possibility because brane intersections typically result in deformations of the underlying branes.  The most general possibility is to use frames, and hence $\Gamma$-matrices that are an arbitrary  $SO(3)$ rotation (depending on $(u,v,z)$) of the frames $(e^2,e^3,e^4)$ in   (\ref{genframes}).   This is a little too challenging to analyze here, and so we make a more physical choice.

If one thinks in terms the IIA theory, we have a system of NS5, NS5' branes and F1 strings.  The former are much heavier than the latter, and so they can be fixed along the coordinate axes while the M2 brane direction can be fibered over the M5 and M5' directions.  This leads to the Ansatz we will use here:
\begin{equation}
\begin{aligned}
e^0 ~=~  & e^{A_0}\, dt \,, \qquad e^1~=~  e^{A_0}\, dy \,,  \qquad e^2 ~=~    e^{A_1}\, \big( dz ~+~B_1  \, d u~+~B_2  \, d v  \big) \,,\\
e^3 ~=~ &  e^{A_2} \, du\,,  \qquad e^4 ~=~   e^{A_3} \, dv  \,, \qquad e^{i+4}~=~ u\,  e^{A_4} \,  \sigma_i  \,, \qquad e^{i+7}  ~=~  v\, e^{A_5} \,  \tilde \sigma_i \,,   \qquad {i = 1,2,3}   \,.
\end{aligned}
 \label{RotInvframes}
\end{equation}
where $A_0, \dots, A_5$  and  $B_1, B_2$ are arbitrary functions of $(z,u,v)$.  We have, for convenience, introduced  factors of $u$ and $v$ into the definitions of the $e^{i+4}$ and $e^{i+7}$ respectively.  Finally, one can also make a  re-parametrization $z \to   \tilde z(z,u,v)$ so as to gauge away $B_2$ (or $B_1$).  Therefore, without loss of generality, one can take:
\begin{equation}
B_2 ~\equiv~   0 \,.
 \label{gaugechoice1}
\end{equation}
We will therefore adopt the frames:
\begin{equation}
\begin{aligned}
e^0 ~=~  & e^{A_0}\, dt \,, \qquad e^1~=~  e^{A_0}\, dy \,,  \qquad e^2 ~=~    e^{A_1}\, \big( dz ~+~B_1  \, d u \big) \,,\\
   e^3 ~=~ &  e^{A_2} \, du\,,  \qquad e^4 ~=~   e^{A_3} \, dv  \,, \qquad e^{i+4}~=~ u\,  e^{A_4} \,  \sigma_i  \,, \qquad e^{i+7}  ~=~  v\, e^{A_5} \,  \tilde \sigma_i \,,   \qquad {i = 1,2,3}   \,,
\end{aligned}
 \label{finalframes}
\end{equation}
and metric:
\begin{equation}
\begin{aligned}
ds_{11}^2 ~=~ e^{2  A_0}\, \big(- dt^2 ~+~ dy^2\big)   
& ~+~ e^{2A_2} \, du^2 ~+~ e^{2A_3} \, dv^2  ~+~ u^2 \,  e^{2A_4} \, d\Omega_3^2    ~+~ v^2 \,  e^{2A_5} \, d{\Omega'}_3^2  \\ &~+~ e^{2  A_1}\, \big( dz ~+~B_1  \, d u  \big)^2     \,.
\end{aligned}
 \label{11metric-final}
\end{equation}
Within this Ansatz there remains the freedom to re-parametrize $z \to   \hat z(z,u)$, and to re-define $u \to \hat u(u)$, $v \to \hat v(v)$.  

It is simpler to make an appropriately invariant Ansatz for the four-form field strength:
\begin{equation}
\begin{aligned}
F^{(4)} ~=& ~    e^0 \wedge e^1 \wedge\, \big (\,b_1 \,  e^2 \wedge e^3 ~+~ b_2 \, e^2 \wedge e^4  ~+~ b_3 \,  e^3 \wedge e^4 \,\big )  \\
&+~\big(\,b_4 \, e^2 + b_5 \, e^3 +  b_6 \, e^4 \,\big) \wedge e^5 \wedge e^6 \wedge e^7   ~+~\big(\,b_7 \, e^2 +  b_8 \, e^3 + b_9 \, e^4 \,\big) \wedge e^8 \wedge e^9 \wedge e^{10} \,,
\end{aligned}
 \label{F4gen}
\end{equation}
where $b_1, \dots, b_9$ are arbitrary functions of $(u,v,z)$.  One will ultimately have to impose the Bianchi identities on $F^{(4)}$.

\subsection{Solving the BPS system}
\label{ss:BPSsystem}

If one uses the fact that $\bar \epsilon \Gamma^\mu \epsilon$ is necessarily the time-like Killing vector $\frac{\partial}{\partial t}$ one finds that the $(u,v,z)$ dependence of the Killing vector is determined by:
\begin{equation}
\epsilon ~=~  e^{\frac{1}{2} A_0 } \, \epsilon_0 \,,  
 \label{epsnorm}
\end{equation}
where $ \epsilon_0$ is independent of $t,y,z,u,v$.  The dependence of the supersymmetries on the sphere coordinates is determined entirely by the representations of $SO(4) \times SO(4)'$, or $(SU(2))^4$: four out of the eight supersymmetries are independent of the sphere angles and four rotate in the vector representation of each $SO(4)$ (or as bi-fundamentals of each pair of $SU(2)$'s).

Using this, the projectors (\ref{projs3}) and the Ansatz  (\ref{11metric-final})  and  (\ref{F4gen}), it is straightforward to solve the hugely over-determined system (\ref{11dgravvar}).

A first pass through this system determines the functions $b_i$ algebraically in terms of the $A_j$ and $B_1$ and the first derivatives of the $A_j$ and $B_1$.  One then eliminates the $b_i$ entirely to arrive at a collection of first-order differential constraints on the $A_j$ and $B_1$.

This collection includes: 
\begin{equation}
\partial_u \big(A_5 - A_3 \big) ~=~  \partial_z \big(A_5 - A_3 \big) ~=~  0 \,, \qquad \partial_v \big(A_4 - A_2 \big) ~=~  \partial_z \big(A_4 - A_2\big) ~=~  0 \,.
 \label{firstreln}
\end{equation}
This means that   $(A_5 - A_3 )$ is only a function of $v$ and $(A_4 - A_2)$ is only a function of $u$.  Remembering that the Ansatz still allows the re-definition $u \to \hat u(u)$, $v \to \hat v(v)$, we can absorb these functional dependences of $(A_5 - A_3 )$  and $(A_4 - A_2)$ into such a coordinate re-definition and assume, without loss of generality, that 
\begin{equation}
A_4 ~=~ A_2   \,, \qquad A_5 ~=~ A_3  \,,
 \label{Areln1}
\end{equation}
which means that the sphere metrics in $ds_{11}^2$ extend  to the metrics of two conformally flat $\IR^4$'s:
\begin{equation}
\begin{aligned}
ds_{11}^2 ~=~ e^{2  A_0}\, \big(- dt^2 ~+~ dy^2\big)   
& ~+~ e^{2A_2} \,\big( du^2~+~ u^2 \,  \, d\Omega_3^2\big)   ~+~ e^{2A_3} \,\big( dv^2~+~ v^2 \,  \, d{\Omega'}_3^2 \big)  \\ &~+~ e^{2  A_1}\, \big( dz ~+~B_1  \, d u  \big)^2     \,.
\end{aligned}
 \label{11metric-reduced1}
\end{equation}

Using (\ref{Areln1}), some of the other first-order equations show that $(A_0 + A_2 + A_3)$ is a constant.   This constant can be taken to be zero by scaling $u$ and $v$, and so we can take: 
\begin{equation}
A_3 ~=~ -(A_0    +  A_2 ) \,.
 \label{Areln2}
\end{equation}

The first order system then gives $\partial_v (A_1 +2 A_2 )= 0$, which means that  $A_1= -2 A_2 + a_1 (z,u)$ for some arbitrary function, $a_1$.  However there is still the freedom to re-define $z \to   \hat z(z,u)$, and so we can take $a_1 \equiv 0$, to arrive at:
\begin{equation}
A_1 ~=~ -2 A_2 \,.
 \label{Areln3}
\end{equation}

We have thus simplified the eleven-dimensional metric to the form:
\begin{equation}
\begin{aligned}
ds_{11}^2 ~=~ e^{2  A_0}\, \bigg[ \,\big(- dt^2 ~+~ dy^2\big)   
& ~+~ e^{2(A_2-A_0)} \,\big( du^2~+~ u^2 \,  \, d\Omega_3^2\big)   ~+~ e^{-2(A_2+ 2A_0)}  \,\big( dv^2~+~ v^2 \,  \, d{\Omega'}_3^2 \big)  \\ &~+~ e^{-2  (A_0 + 2 A_2)}\, \big( dz ~+~B_1  \, d u  \big)^2   \bigg]  \,.
\end{aligned}
 \label{11metric-reduced2}
\end{equation}

There remains one last differential constraint in the first-order system:
\begin{equation}
\partial_z \big( B_1 \, e^{-2  (A_0 + 2 A_2)}\big) ~=~ \partial_u \big( e^{-2  (A_0 + 2 A_2)}\big)  \,.
 \label{diifconstr1}
\end{equation}
This can be solved by introducing a potential, $w(u,v,z)$, with:
\begin{equation}
 B_1 \, e^{-2  (A_0 + 2 A_2)}  ~=~- \partial_u w \,, \qquad   e^{-2  (A_0 + 2 A_2)}~=~  -\partial_z  w \,,
 \label{potentialw1}
\end{equation}
which leads to 
\begin{equation}
 B_1   ~=~(\partial_z  w)^{-1} \,\partial_u w \,, \qquad   e^{-2  (A_0 + 2 A_2)}~=~  -\partial_z  w \,,
 \label{potentialw2}
\end{equation}
and the metric  (\ref{11metric-reduced2}) becomes exactly that of (\ref{11metric-symm}).

One then finds that all the BPS equations are satisfied.  However, one still has to solve the Bianchi conditions on $F^{(4)}$.

\subsection{Solving the Bianchi equations}
\label{ss:Bianchis}

Solving the BPS equations led to expressions for the $b_i$ in terms of the $A_j$ and $B_1$ and their first derivatives.  One thus obtains expressions for the $b_i$ in terms of the $A_j$, the first derivatives of $A_j$ and the first and second derivatives of $w$.  The Bianchi identities thus lead to equations that are third-order in derivatives of $w$. Amazingly enough, these equations can be integrated.

Define:
\begin{equation}
F_1   ~\equiv~ (-\partial_z  w)^{\frac{1}{2}} \,e^{-3   A_0} \,, \qquad  F_2  ~\equiv~ (-\partial_z  w)^{-\frac{1}{2}} \,e^{-3   A_0}~+~  (-\partial_z  w)^{-1} \,  (\partial_u w)^2\,,
 \label{F12defns}
\end{equation}
and then set:
\begin{equation}
H_1   ~\equiv~\cL_v w  ~-~  \partial_z F_1 \,, \qquad H_2   ~\equiv~\cL_u w  ~+~  \partial_z F_2  \,,
 \label{H12defns}
\end{equation}
where $\cL_u$ and $\cL_v$ are the Laplacians on the $\IR^4$'s.

The Bianchi identities can be summarized as 
\begin{equation}
\partial_z H_1   ~=~  \partial_u H_1  ~=~  \partial_z H_2~=~  \partial_v H_2~=~ 0  \,,
 \label{Bianchis1}
\end{equation}
and hence $H_1 = H_1(v)$ and $H_2 = H_2 (u)$.

One should note that (\ref{potentialw1}) only defines $w$ up to the addition of an arbitrary function of $v$, and so we can take $H_1(v) \equiv 0$.  We will simplify life by taking $H_2(u) \equiv 0$.  Having set $H_1 \equiv H_2 \equiv 0$, one can  satisfy (\ref{H12defns}) by introducing a pre-potential, $G_0(u,v,z)$, with:
\begin{equation}
w ~=~  \partial_z G_0 \,, \qquad   F_1   ~=~ \cL_v G_0  \,, \qquad  F_2  ~=~  -\cL_u G_0  \,.
 \label{G0defn}
\end{equation}
From  this and (\ref{F12defns}), one can determine $e^{-3   A_0}$ in terms of $\cL_v G_0$ and $(-\partial_z  w)^{\frac{1}{2}}$.  Substituting this into the second  expression in  (\ref{F12defns}) and using (\ref{G0defn}), one obtains an equation that determines $G_0$:
\begin{equation}
  \cL_v G_0  ~=~  (\partial_z^2  G_0) \,(\cL_u G_0) ~-~  (\partial_u \partial_z  G_0 )^2    \,, 
 \label{MALap}
\end{equation}
which is precisely the spherically symmetric form of  (\ref{master}).

\section{The democracy of M5 and M5' branes}
\label{app:democracy}

The metric \eqref{11frames} and fluxes \eqref{C3gen} given  in Section \ref{ss:metric} appear to be asymmetric between the two $\IR^4$'s, and hence between the M5 and M5' branes.

The purpose of this Appendix is to show that this is a coordinate artifact inherent in the fibration of the M-theory direction. Following a discussion in \cite{Lunin:2007mj}, we will show that one can flip the fibration from the $\vec u$-plane to the  $\vec v$-plane by exchanging the role of $w$ and $z$. In the $\vec u$-plane fibration \eqref{11frames}, $w$ is a function and $z$ is a coordinate. In the $\vec v$-plane fibration we will construct here, $w$ is a coordinate and $z$ is a function appearing in the solution, $z(w,\vec u,\vec v)$.   

It is useful to introduce the notation, familiar from thermodynamics, in which subscripts on parentheses specify the variables that are being held fixed.  For example,  given a function, $F$, and some variables $\eta, \zeta, \xi$, the expression
\begin{equation}
\bigg( \frac{\partial F}{\partial \eta}\bigg)_{\zeta, \xi}
 \label{DiffNotn}
\end{equation}
specifically indicates that the derivative with respect to $\eta$ is being taken while $\zeta$ and $\xi$ are held fixed. 

Consider the complete differential of the function $w(z,\vec u,\vec v)$:
\begin{equation}
d w~=~ \bigg( \frac{\partial w}{\partial z}\bigg)_{\vec u,\vec v} \, dz ~+~ \bigg( \frac{\partial w}{\partial u_i}\bigg)_{z,\vec v} \, du_i ~+~ \bigg( \frac{\partial w}{\partial v_i}\bigg)_{z,\vec u} \, dv_i \,.
 \label{diffw}
\end{equation}
If one holds $w$ fixed, then this must vanish and one then obtains:
\begin{equation}
 \bigg( \frac{\partial z}{\partial u_i}\bigg)_{w,\vec v}  ~=~ - \bigg( \bigg( \frac{\partial w}{\partial z}\bigg)_{\vec u,\vec v}\bigg)^{-1} \,  \bigg( \frac{\partial w}{\partial u_i}\bigg)_{z,\vec v} \,, \qquad  \bigg( \frac{\partial z}{\partial v_i}\bigg)_{w,\vec u}  ~=~ - \bigg( \bigg( \frac{\partial w}{\partial z}\bigg)_{\vec u,\vec v}\bigg)^{-1} \,  \bigg( \frac{\partial w}{\partial v_i}\bigg)_{z,\vec u}\,,
 \label{diffidents1}
\end{equation}
and 
\begin{equation}
 \bigg( \frac{\partial z}{\partial w}\bigg)_{\vec u,\vec v}  ~=~  \bigg( \bigg( \frac{\partial w}{\partial z}\bigg)_{\vec u,\vec v} \bigg)^{-1} \,.
  \label{diffidents2}
\end{equation}
Using this one finds:
\begin{equation}
\begin{aligned}
e^2 & ~=~    (-\partial_z w )^{\frac{1}{2}} \, \Big( dz ~+~(\partial_z w )^{-1}\,  \big (\vec \nabla_{\vec u} \, w \big)  \cdot  d \vec u \Big)  \\
&~=~ - (-\partial_z w )^{-\frac{1}{2}} \, \Big( (\partial_z w ) dz ~+~   \big (\vec \nabla_{\vec u} \, w \big)  \cdot  d \vec u \Big)  ~=~ - (-\partial_z w )^{-\frac{1}{2}} \, \Big( dw  ~-~   \big (\vec \nabla_{\vec v} \, w \big)  \cdot  d \vec v \Big)  \\
&~=~ - \big((-\partial_w z )_{\vec u,\vec v}\big)^{\frac{1}{2}} \, \bigg ( dw  ~+~    \bigg( \bigg( \frac{\partial z}{\partial w }\bigg)_{\vec v,\vec u}\bigg)^{-1}  \bigg(\frac{\partial z}{\partial v_i}\bigg)_{w,\vec v}  \bigg)  d  v_i \bigg)  \\
&~=~ - (-\partial_w z )^{\frac{1}{2}} \, \Big ( dw  ~+~  (\partial_w  z)^{-1} \big (\vec \nabla_{\vec v} \, z \big)  \cdot  d \vec v \Big) \,  \,.
\end{aligned}
 \label{e2forms}
\end{equation}
One also obtains:
\begin{equation}
C^{(3)} ~=~   - e^0 \wedge e^1 \wedge e^2 ~+~ \frac{1}{3!}\, \epsilon_{ijk\ell} \,  \Big(- (\partial_{u_\ell} z) \,  du^i \wedge du^j \wedge du^k ~+~ (\partial_w z )^{-1}\,  (\partial_{v_\ell} z)  \, dv^i \wedge dv^j \wedge dv^k  \Big)  \,.
 \label{C3gen2}
\end{equation}
One therefore finds that by using $w$ as a coordinate and using $z(w, \vec u, \vec v)$ as a function appearing in the metric, the fibration is now over the $\IR^4$ defined by $\vec v$ and the form of $C^{(3)}$ is similarly inverted compared to (\ref{C3gen2}).  

Thus the BPS solution generically requires a non-trivial fibration over one of the $\IR^4$'s but which $\IR^4$ is a matter of a coordinate choice.  We will remain with the formulation in Section  \ref{ss:metric} where the M-theory direction is fibered over $\IR^4 (\vec u)$.

\section{Dualities from the F1-D1 string web to the D2-D4 string web}
\label{app:duality-D2-D4}

In this appendix we describe in detail the dualities we perform to relate the F1-D1 string-web solution constructed in \cite{Lunin:2008tf} to the M2-M5 solutions we construct in Section \ref{sec:GenRes}. 

In order to perform a T-duality along an  isometry direction $x$, we initially have to express the various fields in the following form:
\begin{align}
\label{form}
    ds^2&=G_{xx}(dx+A_{\mu}dx^{\mu})^2+\hat{g}_{\mu \nu}dx^{\mu}dx^{\nu}\,, \nonumber\\
    B_2&=B_{\mu x}dx^{\mu}\wedge (dx+A_{\mu}dx^{\mu})+\hat{B}_2\,,    \\
    C_p&=C_{(p-1)\,x}\wedge (dx+A_{\mu}dx^{\mu})+\hat{C}_p\,, \nonumber 
\end{align} 
where the hatted forms have no leg along $x$. Then, the transformed fields will be given by
\begin{align}
\label{T-duality}
    d\Tilde{s}^2&=G_{xx}^{-1}(dx+B_{\mu x}dx^{\mu})^2+\hat{g}_{\mu \nu}dx^{\mu}dx^{\nu}\,, \nonumber \\e^{2\Tilde{\phi}}&=G_{xx}^{-1}e^{2\phi}\,,  \\
    \Tilde{B}_2&=A_{\mu}dx^{\mu}\wedge dx+\hat{B}_2\,, \nonumber  \\
    \Tilde{C}_p&=\hat{C}_{p-1}\wedge(dx+B_{\mu x}dx^{\mu})+C_{(p)x}\,. \nonumber 
\end{align} 
As for the S-duality, the conventions we use for the S-dual fields of a given Type IIB supergravity solution are the following:
\begin{align}
\label{S-duality}
	\Tilde{g}_{\mu \nu}&=\sqrt{C_0^2+e^{-2\phi}}g_{\mu \nu}\,, \quad  e^{-\Tilde{\phi}}=\frac{e^{-\phi}}{C_0^2+e^{-2\phi}}\,, \quad    \Tilde{C}_0=-\frac{C_0}{C_0^2+e^{-2\phi}}\,, \nonumber\\
    \Tilde{B}_2&=-C_2\,, \quad 
    \Tilde{C}_2=B_2\,, \quad
    \Tilde{C}_4=C_4+B_2\wedge C_2 \,.
\end{align}
Using now \eqref{T-duality} and \eqref{S-duality}, the solution obtained by the duality chain mentioned in the beginning of this section is:
\begin{align}
\label{nexttofinal}
    ds^2&=\frac{\sqrt{\det h}}{h_{11}}(du_2^2+du_3^2)+\frac{1}{\sqrt{\det h}}(-dt^2+dy^2) +\sqrt{\det h}\left(e^{3A}h_{ab}dr^adr^b+ds^2_{\mathbb{R}^4} \right)\,, \nonumber \\
    e^{2\phi}&=\frac{\sqrt{\det h}}{h_{11}}\,, \hspace{10pt} B_2=\frac{h_{12}}{h_{11}}du_2 \wedge du_3\,, \\
    C_3&=e^{3A}h_{1a}dt\wedge dr^a \wedge dy \,, \hspace{10pt} C_5=\frac{1}{h_{11}}dt\wedge du_1 \wedge du_2 \wedge du_3 \wedge dy \,. \nonumber
\end{align}

In order to uplift the solution to M-theory we need to determine the magnetic dual of the $C_5$ field. Even though this is enough for our purposes we will for completeness determine the magnetic dual of the $C_3$ field as well. Our conventions for the democratic formalism are the following: 
\begin{align}
\label{conventions}
    F_p&=dC_{p-1} \hspace{88pt} \text{for } p<3 \,, \nonumber \\
    F_p&=dC_{p-1}+H_3\wedge C_{p-3} \hspace{25pt} \text{for } p \geq 3\,, \nonumber \\
    F_6&=\star F_4 \,, \hspace{20pt} F_8=\star F_2 \,.  \nonumber
\end{align}
For simplicity, we will assume spherical symmetry in the $\mathbb{R}^4$ spanned by $\upsilon_i$ and use hyperspherical coordinates to describe it:
\begin{align}
    \upsilon_3&=\upsilon \cos \phi_1 \,, \nonumber \\
    \upsilon_4&=\upsilon \sin\phi_1 \cos\phi_2  \,, \nonumber \\
    \upsilon_5&=\upsilon \sin \phi_1 \sin \phi_2 \cos \phi_3 \,, \\
    \upsilon_6&=\upsilon \sin \phi_1 \sin \phi_2 \sin \phi_3 \,, \nonumber \\
    ds_{\mathbb{R}^4}^2&=d\upsilon^2+\upsilon^2\left(d\phi_1^2+\sin^2\phi_1\left(d\phi_2^2+\sin^2\phi_2 \,d\phi_3^2\right)\right) \, \nonumber 
\end{align}
Now the metric $h_{ab}$ is a function of $z$, $u_1$ and $\upsilon$.

In order to find the $C_3$ field dual to the $C_5$ of \eqref{nexttofinal} we need to compute $F_6^e=dC_5^e+H_3\wedge C_3^e$~\footnote{The superscripts $e$ and $m$ will be used to denote respectively the electric and magnetic parts of the RR fields.}:
\begin{align}
    dC_5^e&=-\frac{1}{h_{11}^2}\left(\partial_z h_{11}dz + \partial_{\upsilon} h_{11} d\upsilon \right)\wedge dt \wedge du_1 \wedge du_2 \wedge du_3 \wedge dy\,,  \\
    H_3\wedge C_3^e&=\left[ \partial_z \left( \frac{h_{12}}{h_{11}} \right) e^{3A}h_{12}-\partial_{u_1} \left( \frac{h_{12}}{h_{11}} \right) e^{3A}h_{11} \right] dt\wedge du_1 \wedge du_2 \wedge du_3 \wedge dz \wedge dy \nonumber \\
    &-\partial_{\upsilon} \left( \frac{h_{12}}{h_{11}} \right)e^{3A}h_{11} \,dz \wedge dt \wedge d\upsilon \wedge du_2 \wedge du_3  \wedge dy   \\
    &-\partial_{\upsilon} \left( \frac{h_{12}}{h_{11}} \right)e^{3A}h_{12} \, du_1 \wedge dt \wedge d\upsilon \wedge du_2 \wedge du_3 \wedge dy \nonumber \,. 
\end{align}
Summing these two expressions we obtain:
\begin{align}
    F_6^e&=f_1\,dt\wedge dz \wedge du_1 \wedge du_2 \wedge du_3 \wedge dy  \nonumber \\
          &+f_2\, dt \wedge du_1 \wedge du_2 \wedge du_3 \wedge dy \wedge d\upsilon \\
          &-f_3\, dt \wedge dz \wedge du_2 \wedge du_3 \wedge dy \wedge d\upsilon \,, \nonumber 
\end{align}
where the $f_i$ are given by 
\begin{align}
    f_1&=\frac{1}{h_{11}^2}\partial_z h_{11}-\partial_z \left( \frac{h_{12}}{h_{11}} \right)e^{3A}h_{12}+\partial_{u_1}\left( \frac{h_{12}}{h_{11}} \right)e^{3A}h_{11} \,, \nonumber \\
    f_2&=\frac{1}{h_{11}^2}\partial_{\upsilon} h_{11}-\partial_{\upsilon} \left( \frac{h_{12}}{h_{11}} \right)e^{3A}h_{12} \,, \\
    f_3&=\partial_{\upsilon} \left( \frac{h_{12}}{h_{11}} \right) e^{3A}h_{11} \,. \nonumber 
\end{align}
We can now compute $F_4^m$ by $F_4^m=-\star F_6^e$:
\begin{align}
    F_4^m =& -\upsilon^3 h_{11}\left( f_2 h_{11}+f_3 h_{12} \right) dz \wedge d\Omega_3' - \upsilon^3 h_{11} \left( f_2 h_{12} +f_3 h_{22} \right)  du_1 \wedge d\Omega_3' \nonumber\\
    &+ r^3 f_1 h_{11} \det h \, dr \wedge d\Omega_3' \,, 
\end{align}
where $d\Omega_3' = \sin^2\phi_1 \sin \phi_2 \, d\phi_1 \wedge d\phi_2 \wedge d\phi_3$. Using the explicit form of the $f_i$, $F_4^m$ becomes:
\begin{align}
\label{F4m2}
    F_4^m=-&\left(\upsilon^3 \partial_{\upsilon} h_{11} \, dz + \upsilon^3 \partial_{\upsilon} h_{12} \, du_1 \right)\wedge d\Omega_3' \nonumber \\
    &+\upsilon^3\left(h_{22} \partial_z h_{11} -h_{12} \partial_z h_{12} -h_{12} \partial_{u_1} h_{11} + h_{11} \partial_{u_1} h_{12} \right) d\upsilon \wedge d\Omega_3' \,. 
\end{align}
In order to further simplify this expression we substitute \eqref{Kahler} to the first line of \eqref{F4m2} (from now on we ignore $d\Omega_3'$) and get:
\begin{align}
\label{F4a}
    &-\frac{1}{2}\upsilon^3\partial_{\upsilon} \partial_z^2 K\, dz-\frac{1}{2}\upsilon^3 \partial_{\upsilon} \partial_z \partial_{u_1} K \, du_1 = -\frac{1}{2}d (\upsilon^3\partial_{\upsilon} \partial_z K)+\frac{1}{2}\partial_{\upsilon}  (\upsilon^3\partial_{\upsilon} \partial_z K)d\upsilon \\
    =&-\frac{1}{2}d (\upsilon^3\partial_{\upsilon} \partial_z K)+ \frac{\upsilon^3}{2}\partial_z \left( \frac{1}{\upsilon^3} \partial_{\upsilon} (\upsilon^3 \partial_{\upsilon} K)\right)d\upsilon \nonumber = -\frac{1}{2}d (\upsilon^3\partial_{\upsilon} \partial_z K)+ \frac{\upsilon^3}{2}\partial_z \Delta_y K \,d\upsilon \,. \nonumber
\end{align}
Plugging \eqref{Kahler} in the second line of \eqref{F4m2} we get 
\begin{equation}
\label{F4b}
    \frac{\upsilon^3}{4}\left(\partial_{u_1}^2K \partial_z^3 K- 2\partial_z\partial_{u_1}K\partial_z^2\partial_{u_1}K+ \partial_z^2K\partial_z \partial_{u_1}^2K \right)d\upsilon=\upsilon^3 \partial_z (\det h) \,.
\end{equation}
Finally, putting \eqref{F4a} and \eqref{F4b} together we find 
\begin{equation}
    F_4^m=-\frac{1}{2}d (\upsilon^3\partial_{\upsilon} \partial_z K) \wedge d\Omega_3' +\frac{\upsilon^3}{2}\partial_z \left( \Delta_y K + 2 \det h \right)d\upsilon \wedge d\Omega_3' \,. 
\end{equation}
The second term vanishes due to the Monge-Amp\`ere equation \eqref{MongeAmpere} and therefore, from $F_4^m=dC_3^m+H\wedge C_1^m$ and because $C_1=0$, we can easily see that $C_3^m$ is given by 
\begin{equation}
\label{C3m}
    C_3^m=-\frac{1}{2}\upsilon^3\partial_{\upsilon} \partial_z K\, d\Omega_3' \,.
\end{equation}

Let us now find the $C_5$ field dual to the $C_3$ of \eqref{nexttofinal}, for which we need to compute $F_4^e=dC_3^e$:
\begin{align}
    dC_3^e=-&\left[\partial_{u_1}\left(e^{3A}h_{11}\right) -\partial_z \left(e^{3A}h_{12} \right)  \right] dt\wedge du_1 \wedge dz\wedge dy \nonumber \\
    -& \left[ \partial_{\upsilon} \left(e^{3A}h_{11} \right)dz+\partial_{\upsilon} \left(e^{3A}h_{12} \right)du_1 \right]\wedge dt \wedge d\upsilon \wedge dy \,. 
\end{align}
$F_6^m$ will then be given by $F_6^m=\star F_4^e$:
\begin{align}
\label{F6m1}
    F_6^m=&\frac{\upsilon^3 \det h}{h_{11}} \left[h_{12}\partial_{\upsilon} \left(e^{3A}h_{11} \right)-h_{11}\partial_{\upsilon} \left(e^{3A}h_{12} \right)  \right] dz\wedge du_2 \wedge du_3 \wedge d\Omega_3' \nonumber \\
    +& \frac{\upsilon^3 \det h}{h_{11}} \left[h_{22}\partial_{\upsilon} \left(e^{3A}h_{11} \right)-h_{12}\partial_{\upsilon} \left(e^{3A}h_{12} \right)  \right] du_1\wedge du_2 \wedge du_3 \wedge d\Omega_3'  \\
    -& \frac{\upsilon^3 (\det h)^2}{h_{11}} \left[\partial_{u_1} \left(e^{3A}h_{11} \right)-\partial_{z} \left(e^{3A}h_{12} \right)  \right] du_2 \wedge du_3 \wedge d\upsilon \wedge d\Omega_3' \,, \nonumber 
\end{align}
which can be simplified to
\begin{align}
\label{F6m2}
    F_6^m&=\upsilon^3 \left(\frac{h_{12}}{h_{11}}\partial_{\upsilon} h_{11}-\partial_{\upsilon} h_{12} \right)dz \wedge du_2 \wedge du_3 \wedge d\Omega_3' \nonumber \\
    +& \upsilon^3 \left(\frac{h_{12}}{h_{11}}\partial_{\upsilon} h_{12}-\partial_{\upsilon} h_{22} \right) du_1 \wedge du_2 \wedge du_3 \wedge d\Omega_3'  \\
    +& \upsilon^3 \left( \partial_{u_1} (\det h)-\frac{h_{12}}{h_{11}}\partial_z (\det h) \right) d\upsilon \wedge dx_2 \wedge du_3 \wedge d\Omega_3'\,. \nonumber
\end{align}
Using \eqref{Kahler} the first two terms of \eqref{F6m2} can be written as
\begin{equation}
    \frac{1}{2}\frac{h_{12}}{h_{11}}d_{\Sigma}\left(\upsilon^3 \partial_{\upsilon} \partial_z K \right)\wedge du_2 \wedge du_3 \wedge d\Omega_3' -\frac{1}{2}d_{\Sigma} \left( \upsilon^3\partial_{\upsilon} \partial_{_1} K \right) \wedge du_2 \wedge du_3 \wedge d\Omega_3' \,,
\end{equation}
where by $\Sigma$ we denote the two-dimensional space spanned by $z$ and $u_1$. 

We can now compute $C_5^m$ from $F_6^m=dC_5^m+H_3\wedge C_3^m$, where $C_3^m$ is given in \eqref{C3m}: 
\begin{align}
\label{C5m1}
    F_6^m-H\wedge C_3^m=&\left[\frac{h_{12}}{h_{11}}d_{\Sigma} \left(\frac{\upsilon^3}{2}\partial_{\upsilon} \partial_z K \right)+d\left(\frac{h_{12}}{h_{11}}\right)\frac{\upsilon^3}{2}\partial_{\upsilon} \partial_z K \right]\wedge du_2 \wedge du_3 \wedge d\Omega_3' \nonumber \\
    -& d_{\Sigma}\left(\frac{\upsilon^3}{2}\partial_{\upsilon} \partial_{u_1} K \right)\wedge du_2 \wedge du_3 \wedge d\Omega_3' \\
    +& \upsilon^3 \left( \partial_{u_1} (\det h)-\frac{h_{12}}{h_{11}}\partial_z (\det h) \right) d\upsilon \wedge dx_2 \wedge dx_3 \wedge d\Omega_3' \,. \nonumber 
\end{align}
The first two terms of \eqref{C5m1} give
\begin{align}
\label{C5m2}
    d\left(\frac{\upsilon^3}{2}\frac{h_{12}}{h_{11}} \partial_{\upsilon} \partial_z K \right) \wedge du_2 \wedge du_3 \wedge d\Omega_3' -\frac{h_{12}}{h_{11}}\partial_{\upsilon} \left( \frac{\upsilon^3}{2}\partial_{\upsilon} \partial_z K \right)d\upsilon \wedge du_2 \wedge du_3 \wedge d\Omega_3' \nonumber \\
    -d\left(\frac{\upsilon^3}{2}\partial_{\upsilon} \partial_{u_1}K \right)\wedge du_2 \wedge du_3 \wedge d\Omega_3' + \partial_{\upsilon} \left( \frac{\upsilon^3}{2}\partial_{\upsilon} \partial_{u_1}K \right) d\upsilon \wedge du_2 \wedge du_3 \wedge d\Omega_3' \,.
\end{align}
Now if we combine the terms of \eqref{C5m2} that are not total derivatives with the third term of \eqref{C5m1} we get
\begin{equation}
    \frac{\upsilon^3}{2}\partial_{u_1} \left[2\det h + \frac{1}{\upsilon^3}\partial_{\upsilon} \left(\upsilon^3 \partial_{\upsilon} K \right) \right]-\frac{\upsilon^3}{2}\frac{h_{12}}{h_{11}}\partial_z \left[2\det h + \frac{1}{\upsilon^3}\partial_{\upsilon} \left(\upsilon^3 \partial_{\upsilon} K \right) \right]
\end{equation}
and we see that the Monge-Amp\`ere equation \eqref{MongeAmpere} appeared again. Therefore, \eqref{C5m1} reduces to: 
\begin{equation}
    F_6^m-H\wedge C_3^m=d\left(\frac{\upsilon^3}{2}\frac{h_{12}}{h_{11}} \partial_{\upsilon} \partial_z K -\frac{\upsilon^3}{2}\partial_{\upsilon} \partial_{u_1}K \right) \wedge du_2 \wedge du_3 \wedge d\Omega_3' 
\end{equation}
and $C_5^m$ is 
\begin{equation}
    C_5^m=\frac{\upsilon^3}{2}\left( \frac{h_{12}}{h_{11}}\partial_{\upsilon} \partial_z K- \partial_{\upsilon} \partial_{u_1} K \right) du_2 \wedge du_3 \wedge d\Omega_3' \,. 
\end{equation}

To sum up, the final form of the D2-D4 string-web solution is:
\begin{align}
\label{D2-D4final-appendix}
    ds^2&=\frac{1}{\sqrt{\det h}}(-dt^2+dy^2) + \frac{\sqrt{\det h}}{h_{11}}(du_2^2+du_3^2) +\sqrt{\det h}\left(e^{3A}h_{ab}dr^adr^b+ds^2_{\mathbb{R}^4} \right)\,, \nonumber \\
    e^{2\phi}&=\frac{\sqrt{\det h}}{h_{11}}\,, \hspace{10pt} B_2=\frac{h_{12}}{h_{11}}du_2 \wedge du_3\,, \\
    C_3&=e^{3A}h_{1a}dt\wedge dr^a \wedge dy- \frac{\upsilon^3}{2}\partial_{\upsilon} \partial_z K \, d\Omega_3' \,, \nonumber \\ 
    C_5&=\frac{1}{h_{11}}dt\wedge du_1 \wedge du_2 \wedge du_3 \wedge dy + \frac{\upsilon^3}{2}\left( \frac{h_{12}}{h_{11}}\partial_{\upsilon} \partial_z K- \partial_{\upsilon} \partial_{u_1}K \right) du_2 \wedge du_3 \wedge d\Omega_3'\,. \nonumber
\end{align}

\section{The infinite tilted M2-M5 bound state.}
\label{app:tiltedD2}

The Ansatz for the M5-M2 intersections described in Section~\ref{sec:GenRes} is a complicated one. To construct asymptotically-flat solutions, one needs to solve the Monge-Amp\`ere-like equation \eqref{master} with the appropriate boundary conditions. In this Appendix, we consider an alternative approach to construct a simple solution to these equations. We start with a stack of tilted D2-branes, and follow a chain of dualities to obtain a tilted M5-brane solution with M2 flux. We will see how this construction fits the Ansatz of Section~\ref{sec:GenRes}.

A stack of D2 branes is described in Type IIA by the following system:
\begin{align}
    ds^2 &= Z^{-1/2} (-dt^2 + dx_1^2 + dx_2^2) + Z^{1/2}(dx_3^2 + \dots + dx_9^2) \,,
    \\[.5em]
    e^\Phi &= Z^{1/4} \,,
    \\[.5em]
    C_3 &= Z^{-1} dt \wedge dx_1 \wedge dx_2 \,,
\end{align}
where $Z$ is a harmonic function. The branes are smeared along the directions $x_{3,4,5}$, and located at an arbitrary point in the directions $x_{6,7,8,9}$, that we will take to be the center of space. Noting $r$ the distance to the branes in these last four directions, the harmonic function $Z$ takes the form:
\begin{equation}
    Z ~=~ 1+\frac{Q}{r^2} \,,\qquad r^2 \equiv x_6^2 + \dots + x_9^2 \,.
    \label{eq:harmonic}
\end{equation}

We now tilt the system in the $x_{2,3}$ plane by an angle $\theta$. We define the new coordinates $(x_2',x_3')$ by:
\begin{equation}
\begin{aligned}
    x_2 &= x_2' c + x_3' s \,,\\ x_3 &= -x_2' s + x_3' c \,,
\end{aligned}
\end{equation}
where $c \equiv \cos\theta$, $s \equiv \sin\theta$. In the following we will always use the new rotated coordinate and omit the primes. We also introduce the function $W$ as:
\begin{equation}
    W \equiv c^2 Z + s^2 \,.
    \label{eq:defw}
\end{equation}

In the new coordinates, the metric and gauge field of the tilted-brane solution can be expressed as a fibration over the direction $x_3 (\equiv x_3')$:
\begin{align}
    ds^2 &= Z^{-1/2} (-dt^2 + dx_1^2) + Z^{1/2}(dx_4^2 + \dots + dx_9^2) \nonumber\\& \quad+ Z^{-1/2} W \qty(dx_3 - c s (Z-1) W^{-1} dx_2)^2 + Z^{1/2} W^{-1} dx_2^2 \,,
    \\[.5em]
    e^{2\Phi}&= Z^{1/2} \,,
    \\[.5em]
    C_3 &= Z^{-1} s \,dt \wedge dx_1 \wedge \qty(dx_3 - c s (Z-1) W^{-1} dx_2) \nonumber\\&\quad + W^{-1} c \, dt \wedge dx_1 \wedge dx_2 \,.
\end{align}

The goal is now to dualize this solution to a solution of M-theory, by first performing two T-dualities, and then uplifting the solution.

\subsection{Performing two T-dualities} 
\label{ss:t_dualities}

We start by performing two T-dualities, along $x_3$ and $x_4$, using the standard T-duality rules (\ref{form},{\ref{T-duality}).  After the first T-duality along $x_3$, we obtain:
\begin{align}
    ds^2 &=  Z^{-1/2} (-dt^2 + dx_1^2) + Z^{1/2}(dx_4^2 + \dots + dx_9^2) + Z^{1/2} W^{-1} \qty(dx_2^2 + dx_3^2) \,,
    \\[.5em]
    e^{2\Phi} &= W^{-1} Z \,, \qquad\qquad B_2 = -c s (Z-1) W^{-1}\, dx_2 \wedge dx_3 \,,
    \\[.5em]
    C_2 &= Z^{-1} s \, dt \wedge dx_1 \,,\qquad C_4 = W^{-1} c \, dt \wedge dx_1 \wedge dx_2 \wedge dx_3 \,.
\end{align}

This a solution of Type IIB corresponding to a stack of D1-D3 branes, where the D3 branes extend along $(x_1,x_2,x_3)$ and the D1 branes extend along $x_1$.

We then perform the second T-duality, along the $x_4$ direction. Since the solution presents no fibration or B-field in this direction, this is a trivial operation, it yields:
\begin{align}
    ds^2 &=  Z^{-1/2} (-dt^2 + dx_1^2 + dx_4^2) + Z^{1/2}(dx_5^2 + \dots + dx_9^2) + Z^{1/2} W^{-1} \qty(dx_2^2 + dx_3^2) \,, \label{eq:metT4}
    \\[.5em]
    e^{2\Phi} &= Z^{1/2} W^{-1} \,, \qquad\qquad\qquad\qquad B_2 = -c s (Z-1) W^{-1}\, dx_2 \wedge dx_3 \,,
    \\[.5em]
    C_3 &= Z^{-1} s \, dt \wedge dx_1 \wedge dx_4 \,,\qquad C_5 = W^{-1} c \, dt \wedge dx_1 \wedge dx_2 \wedge dx_3 \wedge dx_4 \,.\label{eq:gauge_after_t_duals}
\end{align}

This is a $D2(014)-D4(01234)$ system. Recall that T-dualities preserves the amount of supersymmetries, all the solutions presented here have 16 supersymmetries. 

Note that this solution can be embedded in the ansatz \eqref{D2-D4 final} by making a rotation in the $x_{45}$ plane and relabelling the coordinates. One then identifies
\begin{equation}
\begin{aligned}
    h_{11} = c^2 Z + s^2 \,, \quad h_{22} = s^2 Z + c^2 \\
	h_{12} = c s (Z-1) \,, \qand \det h = Z \,.
\end{aligned}
\end{equation}
We will nonetheless recompute the uplift of the solution to M-theory in this specific instance, as a cross-check to the previous computation, and to identify the solution to the \mname\ equation.

\subsection{The democratic formalism} 
\label{ss:the_democratic_formalism}

To uplift the solution to M-theory, we need to know the full expression of the $C_3$ gauge field in  the democratic formalism. That is to say, we need to determine the magnetic dual of the $C_5$ gauge field of \eqref{eq:gauge_after_t_duals}.

Let us first compute the 6-form field strength $F_6 = dC_5 + dB_2 \wedge C_3$. We have
\begin{align}
    dB_2 \wedge C_3 ~&=~ \qty[- c s \,\partial_l\qty((Z-1) W^{-1}) \,dx_l \wedge dx_2 \wedge dx_3] \wedge \qty[Z^{-1} s\,  dt \wedge dx_1 \wedge dx_4]
    \\
    &=~ c s^2 \, W^{-2} Z^{-1} \qty(\partial_l Z) \, dt \wedge dx_1 \wedge dx_2 \wedge dx_3 \wedge dx_4 \wedge dx_l \,,
\end{align}
and
\begin{align}
    dC_5 ~&=~ c \,\partial_l\qty(W^{-1})\, dx_l \wedge dt \wedge dx_1 \wedge \dots \wedge dx_4
    \\
    &=~ c^3 \, W^{-2} \qty(\partial_l Z) \, dt \wedge dx_1 \wedge dx_2 \wedge dx_3 \wedge dx_4 \wedge dx_l \,.
\end{align}
where there is an implicit summation over $l \in \{5,6,7,8,9\}$, and to compute the derivatives we have used the expression of $W$ in \eqref{eq:defw}.

Summing the two results, we thus obtain
\begin{align}
    F_6 ~&=~ c \, W^{-1} Z^{-1} \qty(\partial_l Z)  \, dt \wedge dx_1 \wedge dx_2 \wedge dx_3 \wedge dx_4 \wedge dx_l
    \\
    &=~ c  Z^{-1} \qty(\partial_l Z) \, e^0 \wedge e^1 \wedge e^2 \wedge e^3  \wedge e^4 \wedge e^l \,,
\end{align}
where $e^i \propto dx^i$ are the natural diagonal frames of the metric \eqref{eq:metT4}.

We can now compute the dual four-form $F_4^{(m)} = - \star F_6$:
\begin{align}
    F_4^{(m)} ~&=~ c \, \frac{\partial_l Z}{Z} \, \frac{\epsilon_{(l-4),abcd}}{4!} e^{4+a} \wedge e^{4+b}\wedge e^{4+c} \wedge e^{4+d} 
    \\
    &=~ c \, \qty(\partial_l Z) \, \frac{\epsilon_{(l-4),abcd}}{4!} dx^{4+a} \wedge dx^{4+b}\wedge dx^{4+c} \wedge dx^{4+d} 
\end{align}
where $\epsilon$ is the rank-5 antisymmetric tensor, the index $l$ still runs between 5 and 9, while the indices $a,b,c,d$ are summed over $ 1,\dots,5$. The exponent ``$(m)$'' of the four-form denotes the magnetic part.

We can further simplify this expression by using the fact that the harmonic function \eqref{eq:harmonic} does not depend on $x_5$, and depends only on the radial direction $r$:
\begin{align}
    F_4^{(m)} ~&=~ - c \, \frac{x_l}{r} (\partial_r Z) \, \frac{\epsilon_{(l-5),abc}}{3!} dx^5 \wedge dx^{5+a} \wedge dx^{5+b}\wedge dx^{5+c}
\end{align}
where now $5$ is excluded from the sum over $l$, $l\in \{6,7,8,9\}$, while the indices $a,b,c$ still run from $1$ to $4$. We then obtain the potential by integrating the field strength:
\begin{align}
    C_3^{(m)} ~&=~ - c  \frac{x_5 x_l}{r} (\partial_r Z) \, \frac{\epsilon_{(l-5),abc}}{3!} dx^{5+a} \wedge dx^{5+b}\wedge dx^{5+c} 
    \\
    &=~ - c \, x_5 \, r^3 (\partial_r Z)\, d\Omega_3' \,.
\end{align}
where $d\Omega_3'$ is the volume form of the unit 3-sphere defined by $x_6^2 + \dots + x_9^2 = 1$.

\subsection{M-theory uplift and matching} 
\label{ss:m_theory_uplift}

We can now uplift the solution to M-theory, calling the new direction $x_{11}$:
\begin{align}
    ds^2 &=  Z^{-1/6} W^{1/3} \qty[Z^{-1/2} (-dt^2 + dx_1^2 + dx_4^2) + Z^{1/2}(dx_5^2 + \dots + dx_9^2)]  \nonumber\\&\quad + Z^{1/3} W^{-2/3} \qty(dx_2^2 + dx_3^2 + dx_{11}^2) \,,
    \\[.5em]
    C_3 &=  Z^{-1} s \, dt \wedge dx_1 \wedge dx_4 - c s (Z-1) W^{-1}\, dx_2 \wedge dx_3 \wedge dx_{11} - c \, x_5 \, r^3 (\partial_r Z) \, d\Omega_3' \,.
\end{align}

The system is now that of M2 branes extending in the directions $(t,x_1,x_4)$, and M5 branes extending in $(t,x_1,x_2,x_3,x_4,x_{11})$. To make contact with the Ansatz of Section~\ref{sec:GenRes}, there remains to apply a rotation in the plane $x_{4,5}$, by the same angle $\theta$ as the first tilt, and to relabel the coordinates to match the notations:
\begin{gather}
    x_4 \to  c u_1 + s z \,,\quad
    x_5 \to -s u_1 + c z \\
    x_1 \to y \,,\quad x_2 \to u_2 \,,\quad x_3 \to u_3 \,,\quad x_{11} \to -u_4  \,,\quad x_{6,7,8,9} \to v_{1,2,3,4} \,.
\end{gather}
where, as previously, $c = \cos\theta, s = \sin\theta$.
The final result is
\begin{align}
    ds^2 &=  W^{1/3} Z^{-2/3} \qty(-dt^2 + dy^2) + W^{4/3} Z^{-2/3} \qty(dz - c s (Z-1) W^{-1} du_1)^2  \nonumber\\&\  + W^{1/3} Z^{1/3} \qty(dv_1^2 + \dots + dv_4^2)  +  W^{-2/3} Z^{1/3} \qty(du_1^2 + du_2^2 + du_3^2 + du_{4}^2) \,, \label{eq:metric_final}
    \\[.5em]
    C_3 &= Z^{-1} s \,dt \wedge dy \wedge ( s\, dz + c\, du_1 ) + c s (Z-1) W^{-1}\, du_2 \wedge du_3 \wedge du_4 + c \,  (s u_1 - c z) \, v^3 (\partial_v Z) \, d\Omega_3' \,, \label{eq:potential_final}
\end{align}
where $v\equiv (v_i v_i)^{1/2}$, the harmonic function is $Z = 1 + Q/v^2$, and $d\Omega_3'$ is the volume form of the unit 3-sphere defined by $\{v^2 = 1\}$. This solution has 4 charges in total: $M2(0y1)$, $M2(0yz)$, $M5(0y1234)$, $M5(0yz234)$. The charges cannot be independently dialed, they are related because the solution preserves 16 supersymmetries. The M2 branes are smeared over the direction $u_{2,3,4}$, and are parallel to the M5 branes in the plane $(z, u_1)$.

Let us now compare this solution with the metric \eqref{11metric}, and with the 3-form potential \eqref{C3gen}, of the Ansatz. To match the metrics, one needs the following identifications:
\begin{equation}
\begin{gathered}
    e^{A_0} ~=~ W^{1/6} Z^{-1/3} \,,\quad (-\partial_z w) ~=~ W \,,\\
    (\partial_{u_1} w) ~=~ c s (Z - 1)  \,.
\end{gathered}
\label{eq:ident}
\end{equation}
As for the potential \eqref{eq:potential_final}, one finds that they can be matched up to a simple gauge transformation, provided we identify:
\begin{equation}
    (\partial_{v_l} w) =  c \,  (s u_1 - c z) \,v_l\, \frac{\partial_v Z}{v} \,.
    \label{eq:dvw}
\end{equation}
The gauge transformation in question is:
\begin{align}
    \delta C_3 ~&=~ - c^2 dt \wedge dy \wedge dz + cs \, dt \wedge dy \wedge du_1 \,.
\end{align}
This confirms that this tilted D2 brane solution can be dualized to the Ansatz considered in this paper. This gives an explicit solution of the \mname\ equation. To see this, first integrate the equations 
\eqref{eq:ident} and \eqref{eq:dvw}, and determine the function $w$:
\begin{equation}
    w ~\equiv~ -z W + cs (Z-1) u_1 \,.
\end{equation}
Then we use \eqref{solfns} and \eqref{reln1} to compute $G_0$:
\begin{equation}
  G_0 = - \frac12 Z (c z - s u_1)^2 - \frac12 (s z + c u_1)^2 + f(v) \,,
\end{equation}
where $f$ satisfies $\cL_v f \equiv v^{-3} \partial_v(v^3 \partial_v f) = Z$. This function satisfies the \mname\ equation \eqref{master}.

\newpage

\begin{adjustwidth}{-1mm}{-1mm} 

\bibliographystyle{utphys}      

\bibliography{references}       

\end{adjustwidth}


\end{document}